\documentclass[11pt]{article}
\usepackage{graphicx}

\begin{document} 

\def\xslash#1{{\rlap{$#1$}/}}
\def \p {\partial}
\def \dd {\psi_{u\bar dg}}
\def \ddp {\psi_{u\bar dgg}}
\def \pq {\psi_{u\bar d\bar uu}}
\def \jpsi {J/\psi}
\def \psip {\psi^\prime}
\def \txi2 {\tilde\xi_2} 
\def \to {\rightarrow}
\def \lrto{\leftrightarrow} 
\def\bfsig{\mbox{\boldmath$\sigma$}}
\def\DT{\mbox{\boldmath$\Delta_T $}}
\def\xit{\mbox{\boldmath$\xi_\perp $}}
\def \jpsi {J/\psi}
\def\bfej{\mbox{\boldmath$\varepsilon$}}
\def \t {\tilde}
\def\epn {\varepsilon}
\def \up {\uparrow}
\def \dn {\downarrow}
\def \da {\dagger}
\def \pn3 {\phi_{u\bar d g}}

\def \p4n {\phi_{u\bar d gg}}

\def \bx {\bar x}
\def \by {\bar y}

% Revising starts on 03.03. 2017 to emphasize the importance. 

\begin{center}
{\Large\bf One-Loop Corrections of Single Spin Asymmetries at Twist-3 in Drell-Yan Processes }
\par\vskip20pt
A.P. Chen$^{1,2}$, J.P. Ma$^{1,2,3}$, G.P. Zhang$^{4}$     \\
{\small {\it
$^1$ Institute of Theoretical Physics, Chinese Academy of Sciences,
P.O. Box 2735,
Beijing 100190, China\\
$^2$ School of Physical Sciences, University of Chinese Academy of Sciences, Beijing 100049, China\\
$^3$ Center for High-Energy Physics, Peking University, Beijing 100871, China\\
$^4$Department of Modern Physics,  University of Science and Technology of China, Hefei, Anhui 230026, China  
}} \\
\end{center}
\vskip 1cm
\begin{abstract}
We study single spin asymmetries at one-loop accuracy in Drell-Yan processes in which one of the initial hadrons is transversely polarized. The spin-dependent part of differential cross-sections can be factorized  with various hadronic matrix elements of twist-2 and twist-3 operators.
These operators can be of even- and odd-chirality.  
In this work, the studied observables of asymmetries are differential cross-sections with different weights. These weights are selected so that the observables are spin-dependent and their virtual corrections are completely determined by 
the quark form factor. 
In the calculations of one-loop corrections we meet collinear divergences in the contributions involving chirality-odd and 
chirality-even operators. We find that all of the divergences can be correctly subtracted. Therefore, our results give 
an explicit example of QCD factorization at one-loop with twist-3 operators, especially, QCD factorization with chirality-odd twist-3 operators.

\vskip 5mm
\noindent
% PACS numbers
\end{abstract}
\vskip 1cm

\par\vskip20pt
\noindent 
{\bf 1. Introduction}
\par 
Single transverse-Spin Asymmetry(SSA) can appear in high energy hadron-hadron collisions in which one of the initial hadrons 
is transversely polarized. For collisions with large momentum transfers one can make predictions by using QCD factorization, 
in which the perturbative- and nonperturbative effects are consistently separated.  
It is well-known that the cross-sections with an unpolarized- or longitudinally polarized hadron can be factorized with hadronic matrix elements of twist-2 operators. These matrix elements are the standard parton 
distribution functions. In the case of SSA the factorization is made with hadronic matrix elements of operators at twist-3, as shown in \cite{EFTE, QiuSt}. SSA is of particular interest in theory and experiment. Nonzero SSA indicates the existence of nonzero 
absorptive part in scattering amplitudes. The matrix elements of twist-3 operators contain more information about inner structure of hadrons 
than those of twist-2 operators. Therefore, it is important to extract them from experiment.         

\par 
In this work we will study SSA in Drell-Yan processes. We will construct two experimental observables, which are differential cross-sections integrated over parts of phase-space with weights. These weights are chosen so that the observables are proportional 
to the transverse-spin. Using them one can extract the spin-dependent part of the full differential cross-section and relevant twist-3 parton distributions. We will study one-loop corrections of the constructed observables. 
\par 
The two observables studied here receive contributions involving various parton distributions. Among them 
twist-3 parton distributions are unknown. It is important to know these twist-3 parton distributions. 
At tree-level, only two twist-3 parton distributions are involved. They are quark-gluon-quark correlations inside hadrons.  
One is of transversely polarized hadron. Its existence implies that partons inside hadrons have nonzero orbital 
angular momenta. Another one is the correlation defined with a chirality-odd operator for an unpolarized hadron. The involved 
contribution is combined with the twist-2 transversity parton distribution, which is not well-known.
In hadronic processes there are usually significant corrections from next-to-leading order.  With our results at one-loop, the twist-3 parton distributions can be extracted from experimental results more accurately than with tree-results. 
At one-loop twist-3 gluon distribution will contribute. Knowing the one-loop correction, it can help to extract 
the twist-3 gluon distribution.   
Currently, the relevant experiment can be perform 
at RHIC and Compass, where transversely polarized proton beam or target are available. 

\par
SSA at tree-level in Drell-Yan processes  has been studied extensively. In \cite{DYSSA,ZM,MaZh,AT,CR,MaZh3} the effect of SSA has been studied in the case where the transverse momentum of the lepton pair is small and approaching to zero. The effect is at order of ${\mathcal O}(\alpha_s^0 )$.
For the case of the large transverse momentum SSA has been studied in \cite{JQVY,KK1,MSZ,KY1, MSS, MSC}, where the effect of SSA 
is at order of ${\mathcal O}(\alpha_s )$.  While calculations beyond tree-level in QCD factorization at twist-2 
are rather standard and many one-loop results exist, there are not many results of one-loop calculation with twist-3 factorization. For Drell-Yan processes there is only one work in \cite{VoYu} where one weighted differential cross-section of SSA  
involving the twist-3 quark-gluon operator of \cite{EFTE,QiuSt} is calculated at one-loop. For Semi-Inclusive DIS different parts of one-loop 
results about SSA can be found in \cite{KVX,DKPV,ShYo}. One-loop study of twist-3 factorization for DIS has been performed in \cite{G2}.     

\par
Recently the complete twist-3 part of the hadronic tensor of Drell-Yan process and of Semi-Inclusive DIS  has been derived 
at the tree-level ${\mathcal O}(\alpha_s^0)$ for the first time in \cite{MaZh3,CMW}, respectively. According to these results one can systematically construct weighted observables of SSA. An interesting finding in these works is that the twist-3 hadronic tensors contain a special part. This special part receives from higher orders of $\alpha_s$ the virtual correction, which is completely  determined by that of the electromagnetic form factor of a quark. The results of higher-order correction of the quark form factor exists in literature and can be easily re-calculated at one-loop. 
In this work, we will construct two weighted differential cross-sections. These two observables 
receive at tree-level contributions only from the special part of the hadronic tensor. Therefore, the one-loop virtual correction to 
the two observables is well-known. 
We then only need to calculate the real corrections to the observables. One can certainly construct observables 
whose tree-level results can receive contributions from other parts of the hadronic tensor besides or except the special part. In this case, 
the virtual correction needs to be calculated and the calculation can be complicated. We leave this for a study in the future.

\par 
In general twist-3 calculations are more complicated than those of twist-2. In the separation of nonperturbative- and perturbative effects the gauge invariance of QCD should not be violated. In \cite{EKT2} it has been shown how the gauge-invariance is maintained. In twist-3 factorization there is a special contribution called 
soft-gluon-pole contribution as shown in \cite{QiuSt}, in which one gluon is with zero momentum entering hard scattering.
It should be noted that the momentum is not exactly zero. In fact the momentum of the gluon is in Glauber region\cite{MSS}. 
The soft-gluon-pole contribution is more difficult to be calculated than others. Interestingly, it is shown in \cite{KoTa,KoTa2,KTY} 
that the soft-gluon-pole contribution at tree-level is related to the corresponding twist-2 contribution at tree-level. This simplifies the calculation of obtaining the soft-gluon-pole contribution. With these progresses twist-3 calculations 
can be done in a relatively straightforward way.       

\par

We will calculate the one-loop correction of the two observables. The contributions to the observables can be divided into two parts. One part contains hadronic 
matrix-elements of chirality-even operators, while another part involves chirality-odd operators. In calculating the chirality-even-
and chirality-odd contributions at one-loop, one will have I.R.- or collinear divergences. The I.R. divergences will be cancelled in the sum of all contributions. The collinear divergences can be correctly factorized into hadronic matrix-elements.      
The final results are finite. Unlike the collinear factorization at twist-2 for DIS- and Drell-Yan processes, where the twist-2 factorization has been proven to hold at all orders, there is no proof of the collinear factorization at twist-3 
at all orders. To show the factorization it is important to perform calculations beyond the tree-level, because of that 
collinear- and I.R.- divergences do not appear at tree-level. They appear at one-loop or higher orders. These divergences are potential sources to violate the factorization.  
Our work presented here gives an explicit example of twist-3 factorization at one-loop. Especially,
it is the first time in the case of the factorization involving chirality-odd operators at one-loop.    
\par      

\par 
Our paper is organized as follows. In Sect. 2 we introduce our notations and derive the tree-level results. In Sect. 3 and Sect. 4
we give the one-loop corrections for the chirality-even- and chirality-odd contributions, respectively. In these sections, 
we also perform the subtraction of the collinear contributions. The collinear singularities will be subtracted into 
various parton distributions. In Sect. 5 we give our final results which are finite. Sect. 6 is our summary.

\par\vskip20pt
\noindent
{\bf 2. Notations and Tree-Level Results}
\par
We consider the Drell-Yan process:
\begin{equation}
  h_A ( P_A, s) + h_B(P_B) \to \gamma^* (q) +X \to  \ell^- (k_1)  + \ell ^+ (k_2)   + X,
\end{equation}
where $h_A$ is a spin-1/2 hadron with the spin-vector $s$ and the spin of $h_B$ is zero or averaged. 
We will use the light-cone coordinate system, in which a
vector $a^\mu$ is expressed as $a^\mu = (a^+, a^-, \vec a_\perp) =
((a^0+a^3)/\sqrt{2}, (a^0-a^3)/\sqrt{2}, a^1, a^2)$.   We introduce two light-cone vectors $l^\mu =(1,0,0,0)$ and $n^\mu =(0,1,0,0)$. Using the two vectors we define two tensors:
\begin{equation}
  g_\perp^{\mu\nu} = g^{\mu\nu} - n^\mu l^\nu - n^\nu l^\mu,
  \ \ \ \ \ \
  \epsilon_\perp^{\mu\nu} =\epsilon^{\alpha\beta\mu\nu}l_\alpha n_\beta, \ \ \ \
  \epsilon^{\alpha\beta\mu\nu} = -\epsilon_{\alpha\beta\mu\nu},  \ \ \ \ \
  \epsilon^{0123}=1.
\end{equation}
With the transverse metric $g_\perp^{\mu\nu}$ we have $a_\perp^\mu = g_\perp^{\mu\nu} a_\nu$ and $a_\perp^2 = -a_\perp\cdot a_\perp 
= (a^1)^2 + (a^2)^2$. 
The momenta of initial hadrons and the spin of $h_A$ in the light-cone coordinate system are: 
\begin{equation}
P_{A}^\mu \approx (P_{A}^+, 0, 0,0),  \quad P_{B}^\mu \approx ( 0, P_{B}^-, 0,0), \quad  s^\mu = s_\perp^\mu =(0,0, s^1, s^2 ), 
\end{equation}
i.e., $h_A$ moves in the $z$-direction with a large momentum. The invariant mass of the observed lepton pair is $Q^2 =q^2=(k_1 +k_2 )^2 $.
The relevant hadronic tensor is defined as:
\begin{equation}
W^{\mu\nu}  = \sum_X \int \frac{d^4 x}{(2\pi)^4} e^{iq \cdot x} \langle h_A (P_A, s), h_B(P_B)  \vert
    \bar q(0) \gamma^\nu q(0) \vert X\rangle \langle X \vert \bar q(x) \gamma^\mu q(x) \vert
     h_B(P_B),h_A (P_A, s)  \rangle.
\end{equation}
We consider the case with $Q^2 \gg \Lambda_{QCD}^2$.  At leading power of $Q^2$, it is well-known that $W^{\mu\nu}$ is factorized with twist-2 operators, which are used 
to define various standard parton distributions, whose definitions can be found in \cite{JSPDF}. At this order  
$W^{\mu\nu}$ does not depend on the transverse spin $s_\perp$. 

\par  

The $s_\perp$-dependence appears at the next-to-leading order of the inverse-power of $Q$. At this order $W^{\mu\nu}$ can be factorized with twist-3 hadronic matrix elements or twist-3 parton distribution functions. We give the definitions of relevant twist-3 matrix elements in the following. For the transversely polarized $h_A$, there are two relevant twist-3 matrix elements, 
called ETQS matrix elements. They are defined as\cite{EFTE,QiuSt}:
\begin{eqnarray}
&& \int \frac{d \lambda_1 d\lambda_2}{4\pi}
e^{ -i\lambda_2 (x_2-x_1) P^+_A -i \lambda_1 x_1 P^+_A  }
 \langle h_A  \vert
           \bar\psi_i (\lambda_1n ) g_s G^{+\mu}(\lambda_2n) \psi_j(0) \vert h_A \rangle
\nonumber\\
  && = \frac{1}{4} \left [ \gamma^ - \right ]_{ji} \tilde s_{\perp}^{\mu} T_F(x_1,x_2)
    + \frac{1}{4} \left [ i \gamma_5 \gamma^- \right ]_{ji} s^\mu_\perp T_\Delta (x_1,x_2) +\cdots,
\label{tw3q}     
\end{eqnarray}
where $\cdots$ denote irrelevant terms. The vector $\tilde s_\perp^\mu$ is defined as $\tilde s_\perp^\mu = \epsilon_\perp^{\mu\nu} s_{\perp\nu}$. In the above and the following, we will suppress the gauge links between 
field operators at different points of the space-time for a short notation. These gauge links are important for making the definitions 
gauge-invariant. The two twist-3 parton distribution functions defined in Eq.(\ref{tw3q}) have the property:
\begin{equation} 
  T_F (x_1,x_2) = T_F(x_2,x_1),\quad T_\Delta (x_1,x_2) = -T_\Delta (x_2,x_1). 
\end{equation}   
One can define another two twist-3 distributions by replacing the field-strength tensor 
operator in Eq.(\ref{tw3q}) with the covariant derivative $D_\perp^\mu$. In addition to them, there are three twist-3 distributions defined with a product of two quark field operators. Two of them 
are given in \cite{EKT}, and one of them is defined in \cite{CMW}. All of these mentioned twist-3 distributions can be expressed with the two defined in Eq.(\ref{tw3q})\cite{CMW,EKT}. Therefore, we will only use $T_{F,\Delta}$ to express our results. We note here that $T_{F,\Delta}$ are defined with chirality-even operators.  
\par 
There are four twist-3 distributions defined only with gluon fields\cite{Ji3G}. One of them can be defined as:   
\begin{eqnarray} 
T^{(f)}_{G} (x_1,x_2) \tilde s^\mu  =  g_s \frac{ i f^{abc} g_{\alpha\beta} }{P^+_A} 
      \int\frac{d y_1 d y_2}{4\pi} 
   e^{-i P^+_A (y_2 (x_2-x_1) + y_1 x_1)}
 \langle h_A\vert G^{a,+\alpha}(y_1 n) G^{b, +\mu}(y_2 n) G^{c,+\beta} (0) \vert h_A \rangle, 
\label{3GT3}      
\end{eqnarray}
The definition of  $T^{(d)}_{G}$ is obtained by replacing $if^{abc}$ 
with $d^{abc}$. Besides these two distributions $T^{(f,d)}_{G}$ other two twist-3 distributions are defined by replacing $g_\perp^{\alpha\beta}$ 
with $\epsilon_\perp^{\alpha\beta}$ in Eq.(\ref{3GT3}). But, the contributions with these two twist-3 distributions 
do not appear in calculations of our work. 
For the matrix elements
with $f^{abc}$  one has:
\begin{eqnarray}
T^{(f)}_{G}(x_1,x_2) = -T^{(f)}_G(-x_2,-x_1), \quad T^{(f)}_G(x_1,x_2) = T^{(f)}_G(x_2,x_1), 
\label{TSY} 
\end{eqnarray}
Similar relations can be derived for distributions defined with $d^{abc}$. We will use $T^{(f,d)}_{G}$ to give our results. The contributions involving these  
twist-3 distributions are in combination with the twist-2 parton distribution functions of $h_B$. 
\par 
There are contributions to $W^{\mu\nu}$ involving hadronic matrix elements defined with chirality-odd operators. These contributions involve the twist-2 transversity distribution of $h_A$ introduced  
in \cite{JaJi}. It is defined as: 
\begin{equation} 
  h_1 (x)  s_\perp^\mu = \int \frac{ d\lambda }{4\pi} e^{-ix \lambda P^+_A} \langle h_A \vert \bar \psi(\lambda n)   \gamma^+ \gamma^\mu_\perp\gamma_5 
   \psi(0) \vert h_A \rangle.   
\end{equation} 
The twist-3 chirality-odd distributions of $h_B$ appear in the contributions.  
For 
the unpolarized hadron $h_B$ we can define 
\begin{equation}  
  T_{F}^{(\sigma) } (y_1,y_2) = - \frac{2 g_s}{d-2} \int\frac{ d\xi_1^+ d\xi_2^+}{4\pi} e^{-i \xi_1^+ y_1 P_B^- -i \xi_2^+(y_2-y_1) 
     P_B^- }
     \langle h_B\vert \bar q(0) \left (i\gamma_{\perp\mu} \gamma^- \right ) G^{-\mu}(\xi_2^+ l)  q(\xi_1^+ l) \vert h_B \rangle  
\label{COTFD}           
\end{equation}  
with $d$ as the dimension of the space-time. Another twist-3 chirality-odd distribution, called $e(x)$, for the unpolarized hadron is defined with the operator 
$\bar\psi \psi$. With equation of motion one can relate $e(x)$ to $T_F^{(\sigma)}$\cite{JKT,BD,ZYL1}. 

\par 
The complete result for the twist-3 contribution of $W^{\mu\nu}$ in the considered case at the leading order of $\alpha_s$ has been derived in \cite{MaZh3}. It is:
\begin{eqnarray}
W^{\mu\nu}  &=& \frac{1 }{2 N_c} \biggr \{ - T_F^{(\sigma)} (y,y) h_1(x) \biggr [ \frac{1}{2} \frac{\partial \delta^2 (q_\perp)}{\partial q_\perp^\rho } \biggr ( g_\perp^{\mu\rho} \tilde s_\perp^\nu +g_\perp^{\nu\rho}\tilde s_\perp^\mu 
     -g_\perp^{\mu\nu} \tilde s_\perp^\rho \biggr ) +  \frac{\delta^2 (q_\perp)}{P_B\cdot q} \biggr ( P_B^\mu \tilde s_\perp^\nu + P_B^\nu\tilde s_\perp^\mu \biggr ) \biggr ]  
\nonumber\\
   && + \bar q(y)  T_F (x,x)  \biggr [ \frac{\delta^2 (q_\perp)}{ P_A\cdot q  }  
\biggr ( P_A^\mu \tilde s^\nu_\perp + P_A^\nu \tilde s_\perp^\mu \biggr ) 
  +  g_\perp^{\mu\nu} \frac{\partial \delta^2 (q_\perp)}{\partial q_\perp^\rho} 
    \tilde s_\perp^\rho \biggr ]\biggr \} + {\mathcal O}(\alpha_s), 
\label{TOW}          
\end{eqnarray}
where $\bar q (y)$ is the anti-quark distribution function of $h_B$. The momentum $q$ of the lepton pair is parameterized as:
\begin{equation} 
   q^\mu = (x P_A^+, y P_B^-, q_\perp^1,q_\perp^2). 
\end{equation}       
Because the result contains 
$\delta^2 (q_\perp)$ and its derivative, the result should be taken as a tensor distribution, i.e., 
the  $U(1)$-gauge invariance should be understood in the sense of integration. By taking any test function ${\mathcal F}(q_\perp)$ one should have from the invariance: 
\begin{equation} 
  \int d^2 q_\perp  {\mathcal F}(q_\perp)W^{\mu\nu} q_{\mu} =0. 
\label{U1}  
\end{equation}     
The result satisfies this equation and hence is gauge-invariant. In $d$-dimensional space-time the result in Eq.(\ref{TOW}) 
remains the same. An interesting observation in deriving the contribution proportional to the derivative of $\delta^2 (q_\perp)$ 
is that the virtual correction to the contribution beyond tree-level is completely determined by the correction of the quark form factor\cite{MaZh3}.     
\par
Beyond tree-level $W^{\mu\nu}$ contain more contributions with tensor structures different than those at tree-level. In principle one can measure various angular distributions of the out-going lepton to extract different components of $W^{\mu\nu}$. However, this requires large statistics in experiment. It is convenient to use weighted differential cross-sections 
to project out particular angular distributions, which are transverse-spin dependent.  In our case, one can construct weighted differential cross-sections to 
extract the spin-dependent parts of $W^{\mu\nu}$. The differential cross-section after the integration of the phase space 
of the lepton pair can be written as: 
\begin{eqnarray}
\frac{ d\sigma} { d Q^2 d^4 q }
     = \frac{1}{2s q^4}  \int d\Gamma_{\ell^+\ell^-} 
                         L_{\mu\nu} W^{\mu\nu} \delta (q^2-Q^2), \quad 
        d\Gamma_{\ell^+\ell^-} =  \frac{ d^3 k_1}{(2\pi)^3 2k _1^0}\frac{ d^3 k_2}{(2\pi)^3 2 k_2^0} (2\pi)^4\delta^4(k_1 -k_2 -q)                 
\label{dsigma}                         
\end{eqnarray}
with $s= 2P_A^+ P_B^-$ and the leptonic tensor:
\begin{equation}
  L^{\mu\nu} = 4 \left ( k_1^\mu k_2^\nu + k_1^\nu k_2^\mu - k_1\cdot k_2 g^{\mu\nu} \right ) . 
\end{equation} 
We take the electric charge of quarks and leptons as $1$ for simplicity. This charge factor can be easily recovered later in our final 
results by multiplying the factor $(4\pi \alpha Q_q)^2$ with $Q_q$ as the electric charge of the quark $q$ in the unit of 
the electric charge of proton.

\par   
In this work we will consider the weighted differential cross-section with the weight function ${\mathcal O}(q,k_1)$  defined as:
\begin{eqnarray}
\frac{ d\sigma\langle {\mathcal O}(q,k_1)   \rangle  } { dx  d Q^2 }
     = \frac{1} {4 Q^4}   \int d y d^2 q_\perp  d\Gamma_{\ell^+\ell^-} 
                                {\mathcal O}(q,k_1)    L_{\mu\nu} W^{\mu\nu} \delta (q^2-Q^2)                        
\end{eqnarray}
with ${\mathcal O}$ as a function of $q$ and $k_1$. It is clear that the weighted differential cross-section with ${\mathcal O}=1$ is the usual one. We will consider two weights named as ${\mathcal O}_{1,2}$. They are:
\begin{equation} 
  {\mathcal O}_1 = \tilde s_\perp \cdot q_\perp, \quad  {\mathcal O}_2 = \tilde s_\perp \cdot k_1 k_1 \cdot q_\perp 
     +\frac{1}{20} ( 2 Q^2 - 7 q_\perp\cdot q_\perp ) \tilde s_\perp \cdot q_\perp.
\label{O12D}        
\end{equation}
At the first look the second term in the weight ${\mathcal O}_2$  takes a strange form. One may only take the first term as a weight. The reason for the choice of ${\mathcal O}_2$ is that the result becomes simpler after the integration of the lepton-pair phase-space:
\begin{eqnarray} 
                \int d\Gamma_{\ell^+\ell^-}   {\mathcal O} _2(q,k_1)   L_{\mu\nu} 
                  = - \frac{A_\Gamma}{2(d^2-1)} Q^4 \biggr ( q_\perp^\mu \tilde s_\perp^\nu + q_\perp^\nu \tilde s_\perp^\mu 
                     + {\mathcal O}(\epsilon) \biggr ), \quad A_\Gamma = \int d\Gamma_{\ell^+\ell^-} 1 .     
\end{eqnarray}
Without the second term in ${\mathcal O}_2$ in Eq.(\ref{O12D}), the contribution at the order ${\mathcal O}(\epsilon)$ in the above will be at the order ${\mathcal O}(\epsilon^0)$. This makes the end-results very lengthy.
It should be emphasized that it is important to measure the differential cross-section weighted with ${\mathcal O}_2$ 
because it receives the contributions from the transversity $h_1$ at tree-level, as shown in the below.
Such a measurement should not be difficult in experiment, although ${\mathcal O}_2$ looks more complicated than 
${\mathcal O}_1$.  
The two weighted differential cross-section can be measured more easily than the full differential cross-section.    
\par 
With the defined weight ${\mathcal O}_{1,2}$ it is straightforward to obtain the results for the corresponding weighted differential cross-section: 
 \begin{eqnarray} 
 \frac{ d\sigma  \langle {\mathcal O}_1 \rangle  }{ d x d Q^2 } 
   &=& -\frac{A_{\Gamma 1}} {x s Q^2}   \frac{1}{8 N_c }  \vert s_\perp \vert^2 \int \frac{d\xi_1 d\xi_2}{\xi_1\xi_2} \delta (1-\xi_1) 
      \delta (1-\xi_2) 
         \biggr [  \frac{1}{2} (d-4) h_1(x_a ) T_F^{(\sigma)} (y_b,y_b) 
\nonumber\\         
      &&     + (d-2) \bar q(y_b) T_F(x_a,x_a) \biggr ] ,
\nonumber\\
   \frac{ d\sigma \langle {\mathcal O}_2 \rangle } { d x  d Q^2 } 
   &=&  - \frac{A_{\Gamma 2}} {x s}  \frac{1 }{8 N_c}\vert s_\perp \vert^2 \int\frac{d\xi_1 d\xi_2}{\xi_1\xi_2} \delta (1-\xi_1) 
      \delta (1-\xi_2)  
       \biggr [ -(d-2) h_1(x_a) T_F^{(\sigma)} (y_b,y_b) 
\nonumber\\       
  &&  +2  \bar q(y_b) T_F(x_a,x_a) \biggr ],                   
\label{TreeC} 
\end{eqnarray}
with 
\begin{eqnarray} 
  &&  x_a=\frac{x}{\xi_1},\quad  y_b = \frac{Q^2}{x  \xi_2 s }, \quad A_{\Gamma 2}= \frac{A_\Gamma}{2 (d^2-1)},\quad  A_{\Gamma 1}= 2 A_\Gamma \frac{d-2}{d-1}. 
\end{eqnarray}  
These weighted differential cross-sections will only receive the contributions from the spin-dependent part of $W^{\mu\nu}$.  
Since our weights are proportional to $q_\perp$, they only receive the contributions from terms in Eq.(\ref{TOW}) proportional to 
the derivative of $\delta^2 (q_\perp)$. It should be noted that there are contributions in which the parton 
from $h_B$ is a quark. These contributions involve the quark distribution function of $h_B$ and $T_F(-x,-x)$ of $h_A$. 
Similar contributions with chirality-odd operators also exist. In this work we will not list these contributions. These contributions can be obtained with the symmetry of charge-conjugation. 

\par 
We will study the one-loop corrections to the weighted differential cross-sections. Since the weights are proportional to $q_\perp$, the virtual correction 
is well-known as mentioned. We only need then to calculate the real corrections. In the calculations we will meet 
I.R- and collinear divergences. At the end these divergences are either cancelled or correctly subtracted. The final results 
of the one-loop corrections are finite. Because of the subtraction, we note that the contribution 
proportional to $d-4=-\epsilon$ in the first line of Eq.(\ref{TreeC}) will give a nonzero contribution at one-loop.

\par\vskip20pt

\begin{figure}[hbt]
\begin{center}
\includegraphics[width=12cm]{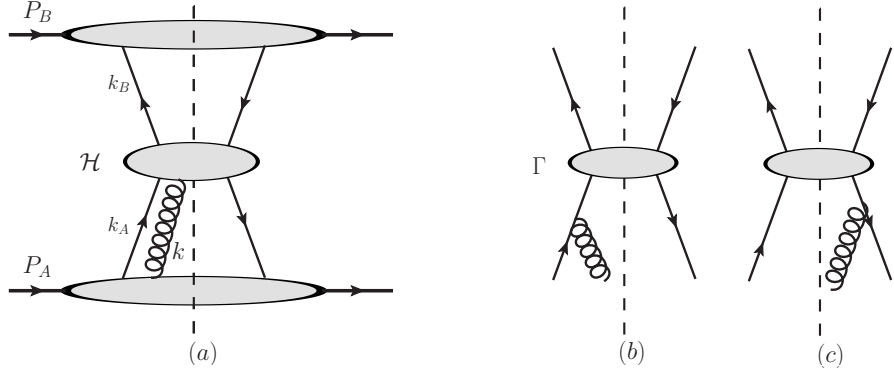}
\end{center}
\caption{Diagrams of one-Loop correction for SSA. The diagrams in (b) and (c) is not included in (a). }  
\label{Feynman-dg2}
\end{figure}

\noindent 
{\bf 3. The One-Loop Correction I} 
\par 
In this section we study one-loop correction involving chirality-even operators and that involving purely gluonic twist-3 operators. In general, we need to calculate diagrams which have the patten illustrated in Fig.1. In these 
diagrams, there is one parton from the hadron $h_B$ participating the hard scattering represented the middle bubble. Fig.1 is for the case that the parton 
from $h_B$ is an anti-quark carrying the momentum $k_B$.  The bubble in the middle denotes those diagrams of the hard scattering. After making the collinear expansion for the antiquark from $h_B$, the contribution like those given in Fig.1
can be written as: 
\begin{eqnarray} 
W^{\mu\nu}\biggr\vert_{Fig.1 } = \int d^4 k_A d^4 k d k_B^-   \frac{1}{2 N_c} \bar q(y_b) 
    \frac{1}{2 N_c} {\rm Tr} \biggr [ \gamma^+ {\mathcal H}^{a,\mu\nu \rho} (k_A,k_B,k)
       {\mathcal M}_g^{a,\rho} (k_A,  k)\biggr ] , 
\label{GS}         
\end{eqnarray} 
with $k_B^\mu =(0,y_b P_B^-,0,0)$ and the quark-gluon correlator 
\begin{eqnarray}             
 {\mathcal M}_g^{a,\rho} (k_A,  k) &=& g_s\int \frac{d^4 \eta_1  d^4\eta_2 }{(2\pi)^8} 
     e^{ i  \eta_1\cdot k_A +i \eta_2 \cdot k } 
       \langle h_A \vert \bar q (0 ) G^{a,\rho} (\eta_2 ) q ( \eta_1 ) \vert h_A \rangle.
\end{eqnarray} 
In the above we have already made some approximations to neglect contributions at twist higher than 3. 

\par 
It is now rather standard to make the collinear expansion related to $h_A$. Here, one should do the 
expansion carefully for obtaining gauge-invariant results. This has been discussed in detail in
\cite{QiuSt,EKT2, EKT}. Since the calculations of twist-3 are now straightforward, we will not give the detail about the calculations. One can find the detail about how to find  gauge-invariant twist-3 contributions in \cite{EKT2,EKT}. 
In the relevant twist-3 contributions, there are contributions in which a gluon with the zero momentum from hadrons
enters the hard scattering. These contributions are called soft-gluon-pole contributions. It is interesting to note that 
there is an elegant way to find such contributions\cite{KoTa,KoTa2}, which we will discuss more in detail in Sub-section 3.2. 
Besides the soft-gluon-pole contributions, there are soft-quark-pole contributions and hard-pole contributions. In the latter 
the momentum component $k^+$ of the gluon is not zero in general. 

\par 
In the real corrections there is always one parton 
in the intermediate state in the hard scattering so that the transverse momentum of the virtual photon becomes nonzero. The square of the transverse momentum is given by:
\begin{equation} 
   q_\perp^2 =-q_\perp \cdot q_\perp = \frac{Q^2}{\xi_2} (1-\xi_1)(1-\xi_2). 
\label{QT}    
\end{equation}    
In this section we will list our results for the hard-pole contribution in the subsection 3.1., for the soft-pole contributions 
in the subsection 3.2, and for the contributions involving purely gluonic twist-3 operators in subsection 3.3.  In the subsection 3.4 we will perform the subtraction for factorizing the collinear contributions into hadronic matrix elements 
to avoid a double counting. After the subtraction the results are finite. 

\par\vskip10pt

\begin{figure}[hbt]
\begin{center}
\includegraphics[width=10cm]{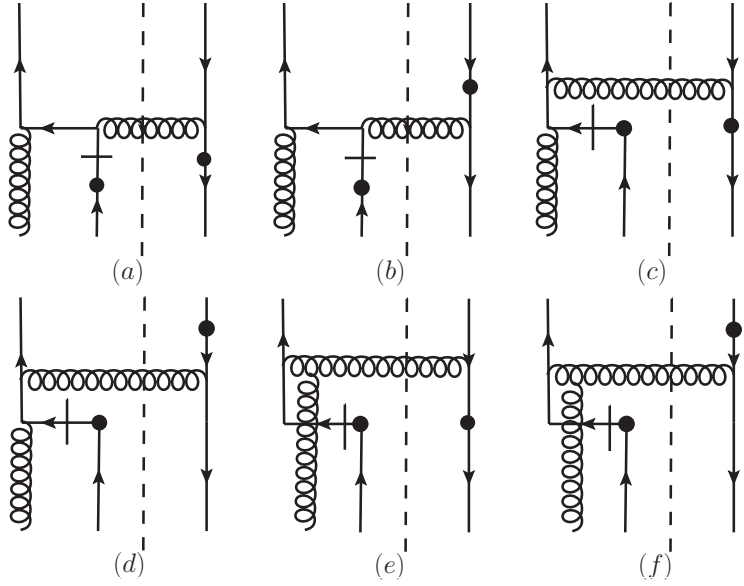}
\end{center}
\caption{Diagrams of the hard-pole contributions. The black dots denote the insertion of the electromagnetic current operators in $W^{\mu\nu}$.  } 
\label{HP1}
\end{figure}

\noindent 
{\bf 3.1. Hard-Pole Contributions}
\par
\begin{figure}[hbt]
\begin{center}
\includegraphics[width=10cm]{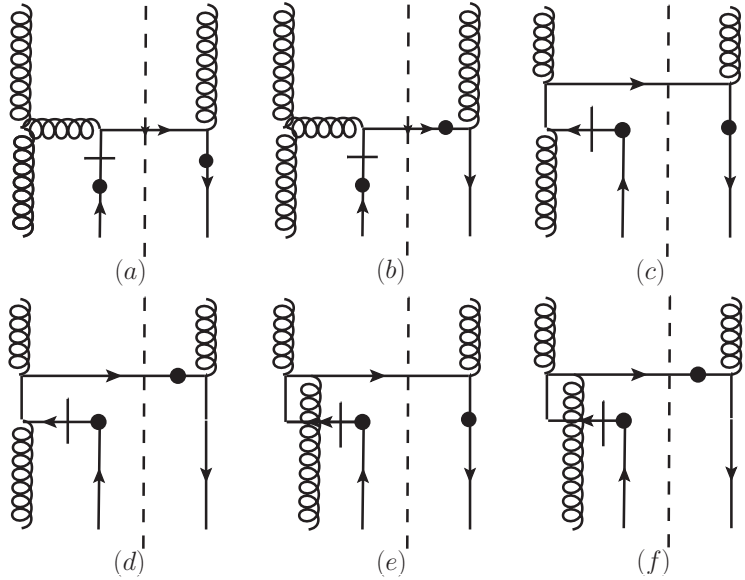}
\end{center}
\caption{Diagrams of the hard-pole contributions with a gluon from $h_B$. } 
\label{HP2}
\end{figure}
\begin{figure}[hbt]
\begin{center}
\includegraphics[width=14cm]{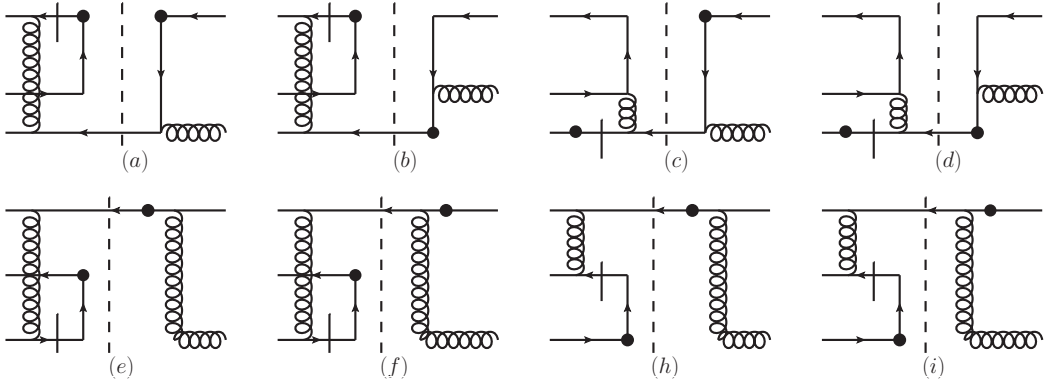}
\end{center}
\caption{The diagrams of the hard pole contributions, where a $q\bar q$-pair from $h_A$ enters the hard scattering.    }
\label{HP3}
\end{figure}

The hard-pole contributions are from diagrams given in Fig.\ref{HP1}, Fig.\ref{HP2} and Fig.\ref{HP3}. These diagrams 
are for the hard scattering represented by ${\mathcal H}$ in Fig.1. In these diagrams, there is a quark propagator 
with a short bar. This is to indicate that we only take the absorptive part of the quark propagator 
in the calculations. The absorptive part is responsible for SSA. 
To calculate our weighted 
differential cross-section, we need to perform the integration over $q_\perp$. The results after the integration contain 
I.R.- and collinear divergences. These divergences come from the momentum region where the momentum of the massless 
parton in the intermediate state is soft or collinear to $P_A$ or to $P_B$. We use the dimensional regularization 
to regularize these soft divergences. In the regularization the dimension of the space-time is $d=4-\epsilon$ and the dimension 
of the transverse space is $2-\epsilon$.  A scale $\mu_c$ related to the soft divergences is introduced.   The calculations are tedious but straightforward.  
\par 
We will use the following notations in our work: 
\begin{eqnarray} 
     F_D =  \biggr ( \frac{ 4\pi \mu_c^2}{Q^2}\biggr )^{\epsilon/2} \frac{1}{\Gamma (1-\epsilon/2)} , 
    \quad L_2 (\xi)=\left (\frac{\ln (1-\xi)}{1-\xi}\right )_+ \quad  L_1 (\xi)=L_2 (\xi) -  \frac{\ln\xi}{1-\xi}.     
\end{eqnarray} 
The $+$-distributions are standard ones. The hard-pole contribution from diagrams in Fig.\ref{HP1} is with an antiquark from $h_B$. The results from these diagrams 
are: 
\begin{eqnarray} 
    \frac{ d\sigma \langle {\mathcal O}_1 \rangle  }{ d x d Q^2}\biggr\vert_{Fig.\ref{HP1}}  &=& \frac{\vert s_\perp \vert^2 A_{\Gamma 1}}{4  x s Q^2}   \frac{\alpha_s} {  4 \pi N_c^2} F_D   \int \frac{ d\xi_1 d\xi_2}{ \xi_1\xi_2} \biggr \{  \bar q(y_b) T_F( x_a, z_a) 
 \biggr [ -2 N_c^2 \left (\frac{2}{\epsilon}\right )^2 \delta (1-\xi_1)\delta(1-\xi_2) 
\nonumber\\ 
   && +\delta (1-\xi_1) 
      (N_c^2 +\xi_2-1)  \frac{2}{\epsilon} \frac{ (1+\xi_2^2) }{(1-\xi_2)_+ } 
 + \delta (1-\xi_2) N_c^2  \frac{2}{\epsilon}\frac{1+\xi_1}{(1-\xi_1)_+}  
   + {\mathcal A}_{1F} (\xi_1,\xi_2)  \biggr ] 
\nonumber\\
    && + \bar q (y_b) T_{\Delta} ( x_a, z_a)  \biggr (  - \delta (1-\xi_2) N_c^2   \frac{2}{\epsilon} 
      + {\mathcal A}_{1\Delta } (\xi_1,\xi_2)  \biggr )  \biggr \}  , 
\nonumber\\
    \frac{ d\sigma \langle {\mathcal O}_2 \rangle  }{ d x d Q^2}\biggr\vert_{Fig.\ref{HP1}}   &=& \frac{\vert s_\perp \vert^2 A_{\Gamma 2}}{4  x s} \frac{\alpha_s } { 4 \pi N_c^2} F_D  \int \frac{ d\xi_1 d\xi_2}{ \xi_1\xi_2} \biggr \{  \bar q(y_b) T_F( x_a, z_a)
 \biggr [ -2 N_c^2 \delta (1-\xi_1)\delta(1-\xi_2)    
\nonumber\\
  &&  \biggr ( \left (\frac{2}{\epsilon}\right )^2 + \frac{2}{\epsilon} \biggr )      + \delta (1-\xi_1) 
      ( N_c^2 +\xi_2-1)   \frac{ 1+\xi_2^2 }{(1-\xi_2)_+ }    \frac{2}{\epsilon} 
      + \delta (1-\xi_2) N_c^2 \frac{ 1+\xi_1}{(1-\xi_1)_+} \frac{2}{\epsilon}
\nonumber\\      
     &&   + {\mathcal A}_{2F} (\xi_1,\xi_2) \biggr ]    
        + \bar q (y_b) T_{\Delta} ( x_a, z_a)    \biggr [-  \delta (1-\xi_2) N_c^2  \frac{2}{\epsilon}  + {\mathcal A}_{2\Delta } (\xi_1,\xi_2) \biggr ]   
  \biggr \} . 
\label{RHP1}    
\end{eqnarray} 
In Eq.(\ref{RHP1}) we list the divergent contributions explicitly. The terms with the functions ${\mathcal A}'s$ 
are finite. In this and the next section we will always give our results in this form. All finite contributions 
will be summed in Sect.5 together and the relevant functions will be given in Appendix.  
The variables $z_a$ is:
\begin{equation} 
  z_a = \frac{x \xi_2}{1-\xi_1 (1-\xi_2) }.
\end{equation}   
We note that the contributions from Fig.\ref{HP1} contain a double-pole in $\epsilon$. 

\par 

\par
The hard-pole contributions from diagrams in Fig.\ref{HP2} are those in which a gluon as a parton from $h_B$. We denote the twist-2 gluon distribution function of $h_B$ as $G(y)$. The contributions from Fig.\ref{HP2} are:   
\begin{eqnarray} 
    \frac{ d\sigma \langle {\mathcal O}_i \rangle  }{ d x d Q^2} \biggr\vert_{Fig.\ref{HP2}}  &=& \frac{\vert s_\perp \vert^2 A_{\Gamma i}}{4  x s Q^{2(2-i)}}  \frac{\alpha_s } { 4 \pi N_c (N_c^2-1) } F_D  \int \frac{ d\xi_1 d\xi_2}{ \xi_1\xi_2} G(y_b)\biggr \{    T_F(x_a,z_a) \biggr [ 2 C_F  {\mathcal B}_{iF} (\xi_1,\xi_2)  
 \nonumber\\
   &&   + \frac{2}{\epsilon} \delta (1-\xi_1) ( (1-\xi_2) N_c^2 -1)  ( 2\xi_2^2 -2 \xi_2 +1)   \biggr ] 
   + 2 C_F T_\Delta (x_a,z_a) {\mathcal B}_{i\Delta } (\xi_1,\xi_2)         
     \biggr \} ,               
\label{RHP2}            
\end{eqnarray} 
for $i=1,2$. The divergent contributions from Fig.\ref{HP2} are the same for $i=1,2$.  
The contributions here contain only a single-pole in $\epsilon$ associated with $T_F$.

\par

\par 
In the contributions from Fig.\ref{HP1} and Fig.\ref{HP2} the momentum fractions $x_1$ for the out-going quark and $x_2$ 
of the in-coming quark, as variables of $T_{F,\Delta} (x_1,x_2)$, are always positive. There are contributions 
in which $x_1$ or $x_2$ is negative. In these contributions there is a quark-antiquark pair from $h_A$ entering the hard scattering. These contributions are from Fig.\ref{HP3}. They are:   
\begin{eqnarray} 
    \frac{ d\sigma \langle {\mathcal O}_i \rangle }{ d x d Q^2}\biggr\vert_{Fig.\ref{HP3}}  &=& \frac{ \vert s_\perp \vert^2 A_{\Gamma i}}{4  x s Q^{2(2-i)} } \frac{\alpha_s } { 4 \pi N_c^2} F_D  \int \frac{ d\xi_1 d\xi_2}{ \xi_1\xi_2} \bar q(y_b) 
\biggr \{
    T_F (-x_\xi, z_a) \biggr (  (1- 2\xi_1) \delta (1-\xi_2) \frac{2}{\epsilon}  
\nonumber\\
   &&  + {\mathcal C}_{iF1}(\xi_1,\xi_2) \biggr )   
 + T_\Delta (-x_\xi,z_a)  \biggr (- \delta (1-\xi_2)\frac{2}{\epsilon} + {\mathcal C}_{i\Delta 1}(\xi_1,\xi_2) \biggr )      
\nonumber\\ 
   && + T_F (-z_a,x_\xi) {\mathcal C}_{iF2} (\xi_1,\xi_2) + T_\Delta  (-z_a,x_\xi) {\mathcal C}_{i \Delta 2} (\xi_1,\xi_2) \biggr \},     
\label{RHP3}      
\end{eqnarray} 
with $x_\xi = x_a -z_a$. 
We notice that the contributions from those 
diagrams in the second row of Fig.\ref{HP3} are finite. The contributions of the first row have the same divergent part 
for $i=1,2$.  
     
\par\vskip30pt

\noindent
{\bf 3.2. Soft-Pole Contributions} 
\par
\begin{figure}[hbt]
\begin{center}
\includegraphics[width=13cm]{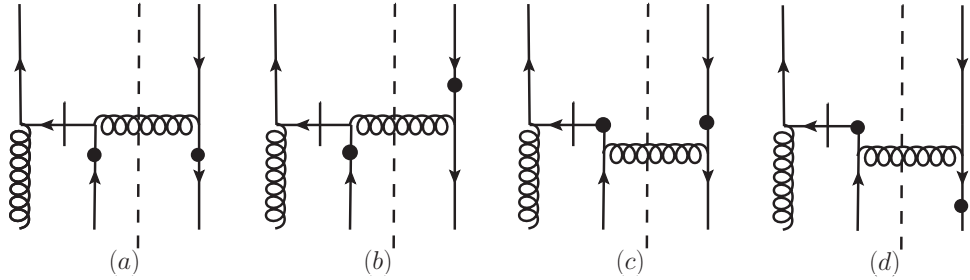}
\end{center}
\caption{Diagrams of the soft-gluon-pole contributions. } 
\label{SP1}
\end{figure}
\begin{figure}[hbt]
\begin{center}
\includegraphics[width=13cm]{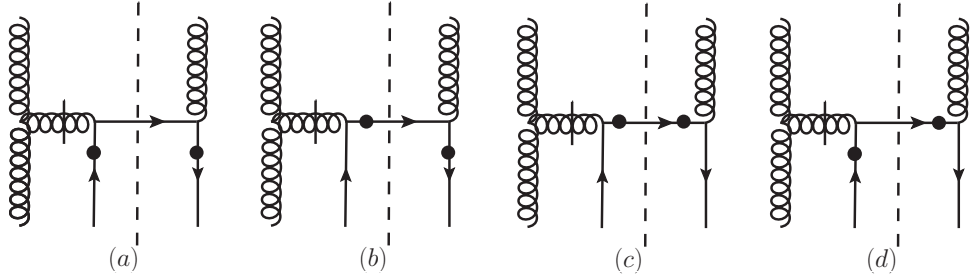}
\end{center}
\caption{Diagrams of the soft-gluon-pole contributions with a gluon from $h_B$. } 
\label{SP2}
\end{figure}
\begin{figure}[hbt]
\begin{center}
\includegraphics[width=13cm]{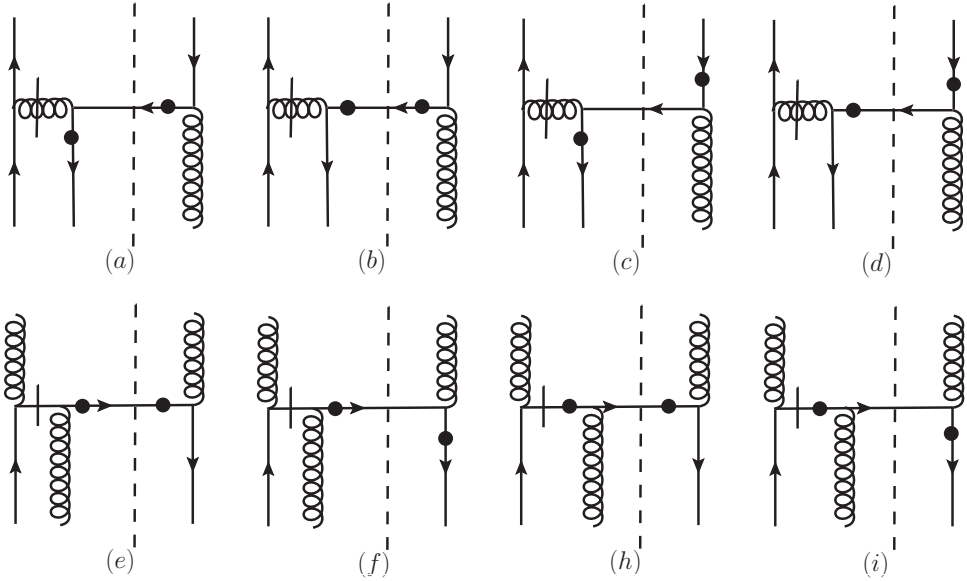}
\end{center}
\caption{Diagrams of the soft-quark-pole contributions as one-loop correction for SSA. } 
\label{SFP1}
\end{figure}

The soft-pole contributions can be soft-gluon-pole- or soft-quark-pole contributions.  
The soft-gluon-pole contributions are from diagrams in Fig.\ref{SP1} and Fig.\ref{SP2}. The soft-quark-pole contributions 
are from the the diagrams in Fig.\ref{SFP1}. The soft-quark-pole contributions can be evaluated directly, while it is complicated to calculate the soft-gluon-pole contributions. However, 
as mentioned, there is an elegant way to obtain the soft-gluon-pole contribution as shown in \cite{KoTa,KoTa2,KTY}. 
In our case, the contributions from Fig.\ref{SP1} can be calculated as discussed in the following. 
\par 
\par

We consider the contribution to the twist-2 part of $W^{\mu\nu}$ from the partonic process 
$q(x_a P_A) +  \bar q (w) \to \gamma^* (q)  +g(k_g)$ at tree-level.  After working  out the color factor, 
the contribution is given by:
\begin{equation} 
  W^{\mu\nu}\biggr\vert_{twist-2} = \frac{N_c^2-1}{8 N_c^2 (2\pi)^4 } \int \frac{dy_b}{y_b} dx_a \bar q(y_b) q(x_a ) S^{\mu\nu}(x_a P_A,y_b P_B ), 
\end{equation}    
where $ q (x_a)$ is the quark distribution function of $h_A$. The quantity $S^{\mu\nu}(x_a P_A,w)$  can be simply calculated from the partonic process. Now the 
soft-pole contribution from Fig.\ref{SP1} to $W^{\mu\nu}$ at twist-3 can be calculated as\cite{KoTa,KoTa2}: 
\begin{eqnarray} 
  W^{\mu\nu} \biggr\vert _{Fig.\ref{SP1}} 
    =  \frac{-\tilde s^\rho }{16 N_c^2 (2\pi)^4} \int \frac{dy_b}{y_b } d x_a  \bar q(y_b ) T_F(x_a,x_a) \biggr  ( \frac{\partial}{\partial w_\perp^\rho} - \frac{k_{B\rho}}{k_B^-} \frac{\partial}{\partial w^+ }
\biggr ) \biggr [ S^{\mu\nu}(x_a P_A,w) \biggr ]\biggr\vert_{w^\rho =k_B^\rho},  
\label{SOFTW}          
\end{eqnarray} 
with $k_B = y_b P_B$.  Similar result can also be derived for the twist-3 contribution from Fig.\ref{SP2}. 

\par 
In the calculation of the soft-gluon-contributions to our weighted differential cross-sections, the obtained results have not only contributions involving $T_F(x_a,x_a)$, 
but also contributions involving the derivative of $T_F(x_a,x_a)$ with respect to $x_a$. These contributions 
after the integration over $q_\perp$ take the form 
\begin{equation} 
  \int_x^1  d x_a (q_\perp^2)^{-\epsilon/2} f(x_a ) \frac{d}{d x_a} T_F (x_a,x_a) = -\int_x^1  d x_a T_F(x_a,x_a) 
       \frac{d}{d x_a} \biggr ( (q_\perp^2)^{-\epsilon/2} f(x_a ) \biggr ),  
\end{equation}              
where $q_\perp^2$ is given in Eq.(\ref{QT}). One can perform integration by part to eliminate these terms with the derivative of 
$T_F(x_a,x_a)$, as shown in the above. Assuming $T_F(1,1)=0$ the contribution from the boundary at $x_a=1$ is zero. 
The contribution from the boundary at $x_a=x$ is also zero in $d$-dimension. If we expand the integral in $\epsilon$ 
and then perform the integration by part, the contribution from the boundary at $x_a=x$ is nonzero and should be taken into 
account. The final results obtained in this way are the same at the considered orders of $\epsilon$, if we perform the integration by part before the expansion in $\epsilon$. We have the contribution from Fig.\ref{SP1}:
\begin{eqnarray} 
    \frac{ d\sigma \langle {\mathcal O}_1\rangle  }{ d x d Q^2} \biggr\vert_{Fig.\ref{SP1}}  &=& \frac{\vert s_\perp \vert^2 A_{\Gamma 1}}{4  x s Q^2}  \frac{\alpha_s } { 4\pi N_c^2} F_D  \int \frac{ d\xi_1 d\xi_2}{ \xi_1\xi_2} \bar q(y_b) T_F (x_a,x_a) 
     \biggr [ 2\delta(1-\xi_1) \delta (1-\xi_2) \biggr ( \biggr (\frac{2}{\epsilon}\biggr )^2 - \frac{2}{\epsilon} \biggr ) 
\nonumber\\
   && -\delta (1-\xi_1) \frac{2}{\epsilon} \frac{(1+\xi_2^2)\xi_2 }{(1-\xi_2)_+} 
     -\delta (1-\xi_2)  
         \frac{2}{\epsilon} \frac{1+\xi_1^2 }{(1-\xi_1)_+}  + {\mathcal D}_1 (\xi_1,\xi_2)  \biggr ] , 
\nonumber\\                  
    \frac{ d\sigma \langle {\mathcal O}_2 \rangle  }{ d x d Q^2}\biggr\vert_{Fig.\ref{SP1}}  &=& \frac{\vert s_\perp \vert^2  A_{\Gamma 2}}{4 x s }  \frac{\alpha_s } { 4 \pi N_c^2} F_D  \int \frac{ d\xi_1 d\xi_2}{ \xi_1\xi_2}  \bar q(y_b) T_F (x_a,x_a)
     \biggr [ 2\delta(1-\xi_1) \delta (1-\xi_2)  \biggr (\frac{2}{\epsilon}\biggr )^2  
\nonumber\\
     &&  -\delta (1-\xi_1) \frac{2}{\epsilon}  \frac{\xi_2 (1+\xi_2^2) }{(1-\xi_2)_+} 
   -\delta (1-\xi_2)\frac{2}{\epsilon} \frac{1+\xi_1^2}{(1-\xi_1)_+} + {\mathcal D}_2 (\xi_1,\xi_2)  \biggr ].
\label{SGP1}       
\end{eqnarray} 
In Eq.(\ref{SGP1}) there are contributions containing  double-poles in $\epsilon$. The contributions with the double poles
will be cancelled by those in the virtual corrections.

\par 
In the contributions from Fig.\ref{SP2} the parton from $h_B$ is a gluon. We have:  
\begin{eqnarray} 
    \frac{ d\sigma \langle {\mathcal O}_i \rangle  }{ d x d Q^2} \biggr\vert_{Fig.\ref{SP2}} &=& \frac{\vert s_\perp \vert^2 A_{\Gamma i}}{4  x s Q^{2(2-i)}}  \frac{\alpha_s N_c } { 4 \pi (N_c^2-1) } F_D  \int \frac{ d\xi_1 d\xi_2}{ \xi_1\xi_2}  G(y_b)  T_F (x_a,x_a) 
\nonumber\\
   && \biggr [ \frac{2}{\epsilon} \delta (1-\xi_1)  ( 2\xi_2^2 -2 \xi_2 +1)\xi_2  + \frac{2 C_F}{N_c^2} {\mathcal E}_i(\xi_1,\xi_2)  \biggr ],      
\label{SGP2}         
\end{eqnarray}  
for $i=1,2$.      

\par 
As shown in \cite{KT1}, 
there are soft-quark-pole contributions, in which one of $x_{1,2}$ in $T_{F,\Delta} (x_1,x_2)$ is zero, i.e., 
a quark or antiquark carrying  zero momentum enters a hard scattering. These contributions are from diagrams given in Fig.\ref{SFP1}, where in the first four diagrams the parton from $h_B$ is an anti-quark, in the remaining diagrams the 
parton is a gluon. The method to calculate these contributions 
is the same as that used for hard-pole contributions in the previous subsection. Interestingly, the results from Fig.\ref{SFP1} 
can be written in a compact form:  
\begin{eqnarray} 
    \frac{ d\sigma \langle {\mathcal O}_i\rangle  }{ d x d Q^2} \biggr\vert_{Fig.7}  &=& \frac{\vert s_\perp \vert^2 A_{\Gamma i}}{4  x s Q^{2(2-i)}}    \frac{\alpha_s } { 4 \pi N_c^2}  \int \frac{ d\xi_1 d\xi_2}{ \xi_1\xi_2}   \biggr ( \bar q(y_b) +\frac{N_c}{N_c^2-1} G (y_b) \biggr )
  \biggr [ T_F(-x_a,0)  {\mathcal F}_{iF}(\xi_1,\xi_2)     
\nonumber\\  
 && 
     + T_\Delta (-x_a,0) {\mathcal F}_{i\Delta }(\xi_1,\xi_2)  \biggr ],       
\label{SQP}        
\end{eqnarray} 
for $i=1,2$. 
The soft-quark-pole contributions here are finite.

\par\vskip10 pt

\noindent 
{\bf 3.3. Gluonic Contribution} 
\par
The gluonic contributions are those in which only gluons from $h_A$ enter the hard scattering. To calculate these contributions it is convenient to use the notation in \cite{Ji3G,KTY,BKTY} for the twist-3 gluonic matrix elements instead of those given
in Eq.(\ref{3GT3}). 
In this notation the matrix element of twist-3 gluonic operator can be parameterized as:
\begin{eqnarray} 
&& \frac{1}{P^+_A } g_s i^3 \int \frac{d\lambda_1}{2\pi }\frac{d\lambda_2}{2\pi} 
   e^{i\lambda_1 x_1 P^+_A + i\lambda_2 (x_2-x_1)P^+_A} \langle h_A  \vert G^{a,+\alpha} 
   (\lambda_1 n)  G^{c,+\gamma}(\lambda_2 n) G^{b,+\beta} (0) \vert h_A  \rangle 
\nonumber\\
   && = \frac{N_c}{(N_c^2-1)(N_c^2-4)} d^{abc} O^{\alpha\beta\gamma}(x_1,x_2) - \frac{i }{N_c(N_c^2-1) } f^{abc} N^{\alpha\beta\gamma}(x_1,x_2), 
\label{NO3G}    
\end{eqnarray} 
where all indices $\alpha,\beta$ and $\gamma$ are transverse.                
With symmetries the two tensors can be decomposed as: 
\begin{eqnarray}
O^{\alpha\beta\gamma}(x_1,x_2) &=& -2 i \biggr [ O(x_1,x_2) g^{\alpha\beta} \tilde s_\perp^\gamma + 
  O(x_2,x_2-x_1) g^{\beta\gamma} \tilde s_\perp^\alpha + O(x_1,x_1-x_2) g^{\gamma\alpha} \tilde s_\perp^\beta \biggr ], 
\nonumber\\
N^{\alpha\beta\gamma}(x_1,x_2) &=& -2 i \biggr [ N(x_1,x_2) g^{\alpha\beta} \tilde s_\perp^\gamma - 
  N(x_2,x_2-x_1) g^{\beta\gamma} \tilde s_\perp^\alpha - N(x_1,x_1-x_2) g^{\gamma\alpha} \tilde s_\perp^\beta \biggr ],         
\end{eqnarray}
with the properties of the function $O$ and $N$ 
\begin{eqnarray} 
    && O(x_1,x_2) =O(x_2,x_1),\quad O(x_1,x_2) = O(-x_1,-x_2), 
\quad  N(x_1,x_2) = N(x_2,x_1),
\nonumber\\ 
   && N(x_1,x_2) = -N(-x_1,-x_2). 
\end{eqnarray}  
These functions are related to those defined in Eq.(\ref{3GT3}) as 
\begin{eqnarray} 
 T_G^{(f)}(x_1,x_2) &=& 2\pi \biggr ( (d-2) N(x_1,x_2) - N(x_2,x_2-x_1) - N(x_1,x_1-x_2) \biggr ), 
\nonumber\\  
   T_G^{(d)}(x_1,x_2) &=& 2\pi \biggr ( (d-2)  O (x_1,x_2) + O(x_2,x_2-x_1) +O (x_1,x_1-x_2) \biggr ).  
\label{NOTG}             
\end{eqnarray}
We will use the relations later to express our final results with $T_G^{(f,d)}$. The obtained results can be conveniently expressed with the combinations: 
\begin{equation}
  T_{G+} (x_1,x_2) = T_G^{(f)}(x_1,x_2) +T_G^{(d)}(x_1,x_2), \quad  T_{G-} (x_1,x_2) = T_G^{(d)}(x_1,x_2) -T_G^{(f)}(x_1,x_2). 
\label{TGPM} 
\end{equation}    
It should be noted that the relations given in Eq.(\ref{NOTG}) depend on $d=4-\epsilon$.  The subtraction of the collinear divergences, as discussed in the next subsection, is determined by the evolution of $T_F (x_a,x_a)$. The gluonic part 
of the evolution derived in the literature is given with $T_{G+}$. Therefore, for the correct subtraction, one should 
re-express the results in terms of $T_{G+}$ and $T_{G-}$ instead of $N$ and $O$. Then the $\epsilon$-dependence 
will deliver an extra contribution.  

\par 
\begin{figure}[hbt]
\begin{center}
\includegraphics[width=13cm]{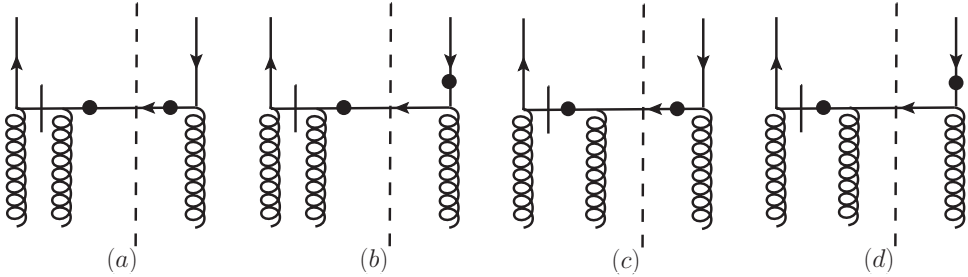}
\end{center}
\caption{Diagrams of the gluonic twist-3 contributions. } 
\label{SGGP1}
\end{figure}
\par 
With the notation in Eq.(\ref{NO3G})  there are only four diagrams giving the gluonic contributions. These diagrams are given in Fig.\ref{SGGP1}. The Bose-symmetry between the three gluons is taken into account with the notation for the twist-3 gluonic matrix elements. The gluonic contributions are of soft-pole contributions, in which one gluon carries zero momentum. 
One can use the method in \cite{KoTa,KoTa2,KTY} to calculate these contributions in a similar way as explained in the previous subsection. Again, in these contributions we will have terms with the derivative on $O$ and $N$. These terms can be 
eliminated with integration by part as discussed before.  We have the results from Fig.\ref{SGGP1}: 
\begin{eqnarray} 
    \frac{ d\sigma \langle {\mathcal O}_i\rangle  }{ d x d Q^2}\biggr\vert_{Fig.\ref{SGGP1}}  &=& \frac{\pi \vert s_\perp \vert^2 A_{\Gamma i}  }{2  x s Q^{2(2-i)} }  \frac{\alpha_s } { 4 \pi N_c } F_D \int \frac{ d\xi_1 d\xi_2}{ \xi_1\xi_2 x_a }  \bar q(y_b)    \biggr \{  \biggr [ -\frac{2}{\epsilon} \delta (1-\xi_2)  2 (2\xi_1^2 -2 \xi_1 +1)    \biggr ] 
 \nonumber\\
   &&    
    \biggr ( O(x_a,x_a) + N(x_a,x_a) + O(x_a, 0) - N(x_a, 0 ) \biggr )  + \frac{1}{N_c} \biggr ( O(x_a,x_a)  
\nonumber\\  
   &&+ N(x_a,x_a) \biggr ) {\mathcal G}_{i+}(\xi_1,\xi_2) 
      + \frac{1}{N_c} \biggr ( O(x_a,0) - N(x_a,0) \biggr ) {\mathcal G}_{i-}(\xi_1,\xi_2)    \biggr \}  ,           
\label{R3G}                    
\end{eqnarray}      
With this result, we have the complete real chirality-even corrections. They are the sum of those results given in Eq.(\ref{RHP1}, \ref{RHP2}, \ref{RHP3}, \ref{SGP1}, \ref{SGP2}, \ref{SQP}, \ref{R3G}).  

\par\vskip10pt

\par\vskip20pt
\noindent 
{\bf 3.4. The Virtual Corrections and  the Subtraction of the Chiral-Even Contributions} 

\par  
As mentioned, the virtual correction to the contributions with the derivative of $\delta^2 (q_\perp)$ in $W^{\mu\nu}$ is determined by the 
quark form factor as observed in \cite{MaZh3}. We will call these contributions as the derivative contributions. 
For self-consistence we explain here in detail how the derivative contributions 
at tree-level appears and hence the observation is made.

\par 
\begin{figure}[hbt]
\begin{center}
\includegraphics[width=5cm]{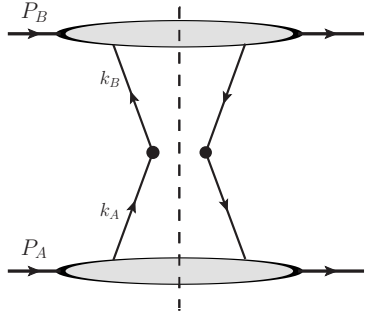}
\end{center}
\caption{A tree-level diagram for the contribution to $W^{\mu\nu}$. Lines for gluon exchanges between quark lines and bubbles are implied(See discussions in text). } 
\label{TreeV}
\end{figure}\par  
The tree-level contributions to $W^{\mu\nu}$ are from Fig.\ref{TreeV}, where there are gluon exchanges between the upper(lower)-bubble 
and the quark lines from the lower(upper)-bubble. At the leading power, we can neglect  all transverse- and $+$-components of 
momenta of gluons exchanged from the upper bubble. At this order, these gluons are polarized in the $-$-direction. The contributions from the exchange of such gluons can be summed into gauge links along the $-$-direction. This results in that the relevant part from the upper 
bubble in the contributions at leading power is represented by the anti-quark parton distribution function of $h_B$, i.e., $\bar q(y_b)$. 
With these approximations the tree-level contribution from Fig.\ref{TreeV} with gluon exchanges is given by
\begin{eqnarray} 
W^{\mu\nu}\biggr\vert_{Tree} &=& \sum_{i=0,j=0} \int d^4 k_A \biggr (\prod_{m=1}^i d^4 k_m \biggr )  \biggr ( \prod_{n=1}^j d^4 \tilde k_n \biggr )  \bar q(y_b) 
\nonumber\\
  &&  \biggr [ \delta^2 (q_\perp -k_{A\perp} - \sum_{m=1}^i 
k_{m \perp} )  {\mathcal H}^{\mu\nu \alpha_1 \cdots \alpha_i \beta_1  \cdots \beta_j} (k_A, \{k_i\}, \{\tilde k_j\})  \biggr ] 
\nonumber\\
   && \int \frac {d^4 \xi}{(2\pi)^4} e^{ik_A\cdot \xi} \biggr ( \prod_{m=1}^i   \frac{d^4\xi_m }{ (2\pi)^4} e^{i k_m\cdot \xi_m } \biggr ) 
    \biggr ( \prod_{n=1}^j   \frac{d^4\eta_n }{ (2\pi)^4} e^{-i \tilde k_n \cdot \eta_n } \biggr ) 
    \langle h_A \vert \bar \psi (0) G^{\beta_1}(\eta_1) \cdots G^{\beta_j}(\eta_j) 
\nonumber\\    
   &&  G^{\alpha_1}(\xi_1) 
    \cdots G^{\alpha_i} (\xi_i) \psi (\xi) \vert h_A\rangle, 
\label{TR9}     
\end{eqnarray}        
where there are exchanges of $i$ gluons in the left part of Fig.{\ref{TreeV} and exchanges of $j$ gluons in the right part.
The gluon fields in the matrix elements and momenta of partons from the lower bubble scale 
like $(1,\lambda^2, \lambda, \lambda)$ with $\lambda\sim \Lambda_{QCD}/Q$. For simplicity we omit 
the color indices in Eq.(\ref{TR9}). In our case we can always neglect the $-$ components of gluon momenta and $k_A^-$ in ${\mathcal H}$. This allows 
to perform the integrations over the $-$-components of momenta and those of space-time components in $+$-directions. 
The contributions with $G^-$ can also be neglected. 

\par    
To find the contributions at twist-3 we need to perform a collinear expansion in which 
we expand the $[ \cdots ] $ in the second line of Eq.(\ref{TR9}) in the transverse momenta. We notice that the twist-2 contributions are obtained by taking the leading order in the expansion and taking all gauge fields 
as $G^+$'s. After summing the contributions of the exchanged gluons into gauge links in the standard way, the twist-2 contributions
are determined by the quark-photon-quark 
vertex at tree-level.   
In the expansion of the $[ \cdots ] $ in transverse momenta, one should also expand the $\delta$-function: 
\begin{equation} 
  \delta^2 (q_\perp -k_{A\perp} - \sum_{m=1}^i  
k_{i\perp} ) = \delta^2 (q_\perp) -\biggr ( k_{A\perp}^\mu +\sum_{m=1}^i  k_{i\perp}^\mu \biggr )  \frac {\partial}{\partial q_\perp^\mu} \delta^2 (q_\perp) +\cdots.  
\label{DDE} 
\end{equation}
In the expansion, the first term gives the contributions starting at order of twist-2, while the leading contribution 
from the second term is at twist-3. 
It is just the second term which gives the derivative contribution of $W^{\mu\nu}$ at tree-level in Eq.(\ref{TOW}). The contribution at twist-3 from this term is then obtained by taking all gauge fields as $G^+$'s and 
neglecting all transverse parton-momenta in ${\mathcal H}$. The calculation is exactly the same as the calculation of twist-2 contributions. The exchange of $G^+$-gluons can be summed with gauge links along the $+$-direction. 
The transverse momenta of partons in the second term can be converted as transverse derivatives acting on parton fields, the final result 
is then expressed with the correlation function: 
\begin{equation} 
  -i q_\partial' (x) \tilde s_\perp^\mu =   \int \frac{d\lambda}{ 4\pi } e^{ i x\lambda  P^+_A} \langle h_A ( P_A,s_\perp)  \vert \bar \psi(0 ) {\mathcal L}_n(\lambda n)  \gamma^+ \partial_\perp^{\mu}   
  \left ( {\mathcal L^\dagger }_n  \psi \right ) (\lambda n ) \vert h_A (P_A,s_\perp) \rangle,  
\label{QPP}           
\end{equation} 
where ${\mathcal L}_n$ is the gauge link in the $+$-direction pointing to the past. A detailed derivation from the second term in Eq.(\ref{DDE}) to the derivative contributions in Eq.(\ref{TOW}) can be found in \cite{MaZh3}. It is shown that $q_\partial'(x)$ is related to $T_F(x,x)$ in \cite{MaZh3,CMW}. After summing the contributions of exchanged gluons emitted from bubbles into gauge links, the derivative contribution is determined by the quark-photon-quark 
vertex, i.e.,  
the quark form factor at tree-level. 
\par 
From the above discussion, it is clear that the derivative contributions are evaluated exactly as the calculation 
of twist-2 contributions except that we have here the correlation function in Eq.(\ref{QPP}) instead of the twist-2 quark 
distribution of $h_A$. 
In Eq.(\ref{TR9})  ${\mathcal H}$ are contributions of tree-level diagrams. For the case that ${\mathcal H}$
contain exchanges of virtual gluons, one can perform the same procedure for the contribution with the second term 
in Eq.(\ref{DDE}). After summing the contributions of exchanged gluons emitted from bubbles into gauge links, 
the derivative contribution is then determined by the 
quark form factor containing exchanges of virtual gluons. 
In the case that there are exchanges of real gluons, i.e., the gluons crossing the cut in Fig.\ref{TreeV}, the 
$\delta$-function in Eq.(\ref{TR9}) is integrated out and the derivative contribution is absent. 
This leads to the observation that the virtual correction
to the derivative contribution is determined by the quark form factor. The same conclusion can also be made for SIDIS. 
The one-loop calculation of the virtual correction involving $T_F$  
for Drell-Yan processe in \cite{VoYu} and  for SIDIS in \cite{KVX} verifies our conclusion explicitly.  The above discussion is for the contribution 
involving chiral-even distributions. The same also holds for the derivative contribution involving chiral-odd distributions.

\par 
The one-loop result of the form factor is well-known. 
 Therefore, we have for the derivative contributions 
of $W^{\mu\nu}$ up to one-loop:   
\begin{eqnarray}
W^{\mu\nu} \biggr\vert_{vir.}   &=& \frac{1 }{4 N_c} \biggr \{ - T_F^{(\sigma)} (y,y) h_1(x) \biggr ( g_\perp^{\mu\rho} \tilde s_\perp^\nu +g_\perp^{\nu\rho}\tilde s_\perp^\mu 
     -g_\perp^{\mu\nu} \tilde s_\perp^\rho \biggr ) + 2 \bar q(y)  T_F (x,x) g_\perp^{\mu\nu} \tilde s_\perp^\rho \biggr \} 
     \frac{\partial \delta^2 (q_\perp)}{\partial q_\perp^\rho }       
\nonumber\\
   &&  \biggr \{ 1 + \frac{\alpha_s C_F}{2\pi} F_D 
    \biggr [  -2 \left(\frac{2}{\epsilon}\right )^2   -3 \biggr (\frac{2}{\epsilon} \biggr )    -8 +\pi^2 \biggr ] +{\mathcal O}(\alpha_s^2 ) \biggr \} 
     +\cdots ,  
\label{WVIR}          
\end{eqnarray}
where $\cdots$ stand for those non-derivative terms. It is noted 
that the one-loop corrections of external legs are included so that the correction does not explicitly depend on the renormalization scale $\mu$ because of the conservation of the electromagnetic current. Including the virtual corrections, the tree-level results in Eq.(\ref{TreeC}) 
are modified by replacement in Eq.(\ref{TreeC}): 
\begin{equation} 
  1 \to \biggr \{ 1 + \frac{\alpha_s C_F}{2\pi} F_D 
    \biggr [  -2 \left(\frac{2}{\epsilon}\right )^2   -3 \biggr (\frac{2}{\epsilon} \biggr )  -8 +\pi^2 \biggr ]   \biggr \}.
\label{VREP}      
\end{equation}     
We note here that the virtual corrections contain a double-pole contribution in $\epsilon$.       

\par
From the results in previous subsections we now add all divergent one-loop chirality-even corrections together. We find the divergent part which can be written as:
 \begin{eqnarray} 
  \frac{ d\sigma \langle {\mathcal O}_1 \rangle  }{ d x d Q^2 } \biggr\vert_{div.}  
   &=& \frac {  \vert s_\perp \vert^2 A_{\Gamma 1}} {4 x s Q^2}  \frac{\alpha_s    }{2 \pi N_c  }     
     F_D   \int \frac{d\xi_1 d\xi_2}{\xi_1\xi_2} \frac{2}{\epsilon}  \biggr \{ \delta (1-\xi_1) T_F (x_a,x_a)      
\nonumber\\
     && \biggr [ \frac{1}{2} (2\xi_2^2 -2\xi_2 +1) G(y_b) +  P_{qq} (\xi_2)  \bar q(y_b) \biggr ] 
      + \delta (1-\xi_2)  \bar q(y_b)
  \biggr [   P_{qq}(\xi_1) T_F(x_a,x_a)
\nonumber\\   
      &&   + \frac{N_c}{2(1-\xi_1)} \biggr ( (1+\xi_1) T_F(x_a,x)- (1+\xi_1^2) T_F(x_a,x_a) \biggr )        
  - N_c \delta (1-\xi_1) T_F(x,x) 
\nonumber\\  
    && - \frac{N_c}{2} T_{\Delta}( x_a,x) 
       +\frac{1}{2 N_c} (1-2\xi_1) T_F(x-x_a,x) 
   -\frac{1}{2 N_c} T_{\Delta}(x-x_a,x)  
\nonumber\\   
   && -\frac{1}{ 2 x_a}   (2\xi_1^2 -2 \xi_1 +1) T_{G+}(x_a,x_a)  
          \biggr] \biggr \}, 
\nonumber\\
   \frac{ d\sigma \langle {\mathcal O}_2 \rangle }{ d x d Q^2 } \biggr\vert_{div.} &=& \frac{A_{\Gamma2} Q^2} {A_{\Gamma_1}} 
   \frac{ d\sigma \langle {\mathcal O}_1 \rangle  }{ d x d Q^2 } \biggr\vert_{div.},            
\end{eqnarray}
with 
\begin{equation} 
  P_{qq} (z) = C_F   \left [ \frac{1+z^2}{(1-z)_+} + \frac{3}{2} \delta (1-z) \right ]. 
\end{equation} 
We note that the double-pole terms are cancelled. The remaining divergent contributions are with a single-pole 
in $\epsilon$. The divergence in the sum is from the momentum region where the parton in the intermediate states
or the exchanged gluon in the virtual correction  
is collinear to $h_A$ or $h_B$. There are contributions from the soft gluon in the virtual correction and in the 
intermediate states. These contributions are proportional to $\delta(1-\xi_1) \delta (1-\xi_2)$.      

\par 
It should be noted that the contributions from the momentum region, where the parton in the intermediate states 
is collinear to $h_A$ or $h_B$, are in fact already included in the hadronic matrix elements of the tree-level 
results given in Eq.(\ref{TreeC}). To avoid a double counting we should consistently subtract the collinear 
contributions in the one-loop correction.        

\par 
We make replacement  in the tree-level results in Eq.(\ref{TreeC}): 
\begin{equation} 
  T_F(x,x) \to  T_F(x,x) - \Delta T_F(x,x), \quad \bar q(y)  \to  \bar q(y)  - \Delta \bar q(y). 
\end{equation}
With the replacement in the tree-level results we have the following quantities at the one-loop accuracy: 
 \begin{eqnarray} 
  \Delta  \frac{  d\sigma \langle {\mathcal O}_1 \rangle  }{ d x d Q^2 } 
   &=& \frac{A_{\Gamma 1}(2-\epsilon)\vert s_\perp \vert^2} {8 x s Q^2 N_c}      \int \frac{d\xi_1 d\xi_2}{\xi_1\xi_2} \delta (1-\xi_1) 
      \delta (1-\xi_2) 
   \biggr (  \bar q(y_b) \Delta T_F(x_a,x_a) + \Delta \bar q(y_b) T_F(x_a,x_a) \biggr ) ,     
\nonumber\\
  \Delta \frac{  d\sigma \langle {\mathcal O}_2 \rangle } { d x  d Q^2 } 
   &=&   \frac{A_{\Gamma 2}\vert s_\perp \vert^2} {4 x s N_c}  \int\frac{d\xi_1 d\xi_2}{\xi_1\xi_2} \delta (1-\xi_1) 
      \delta (1-\xi_2)     
    \biggr ( \bar q(y_b) \Delta T_F(x_a,x_a) + \Delta \bar q(y_b) T_F(x_a,x_a) \biggr ).                  
\label{SUBCE} 
\end{eqnarray}
For the subtraction we should add the above quantities to the calculated one-loop corrections, where $\Delta T_F$ and $\Delta \bar q$ are specified in the following.  We have used the dimensional regularization for U.V.-, I.R.- and collinear divergence. 
With the dimensional regularization 
$\Delta T_F$ and $\Delta \bar q$ are determined by the evolution of the renormalization scale $\mu$, respectively. 
The evolution of $T_F(x,x)$ can be found in \cite{KangQiu1, BMP, ZYL, MW, SchZh, KangQiu2}. We have then:  
\begin{eqnarray} 
\Delta T_F(x,x) &=& \frac{\alpha_s}{2\pi} \left ( -\frac{2}{\epsilon_c} + \ln \frac{e^\gamma \mu^2}{4\pi\mu_c^2 }\right ) 
 \biggr \{-N_c T_F(x,x) +   
    \int_x^1  \frac{dz}{z}   \biggr  [ P_{qq} (z) T_F(\xi,\xi)  + \frac{N_c}{2} \biggr ( T_{\Delta}(x,\xi)  
\nonumber\\    
   &&  +  \frac{ (1+z) T_F(x,\xi) -(1+z^2) T_F(\xi,\xi) }{1-z} \biggr ) 
+ \frac{1}{2N_c} \biggr (  (1-2z) T_{F}(x,x-\xi)
\nonumber\\ 
    &&   + T_{\Delta} (x,x-\xi ) \biggr )  
  -\frac{1}{2}   \frac{ (1-z)^2 + z^2 }{\xi}    
     T_{G+}(\xi,\xi)  \biggr ]  \biggr \} 
\nonumber\\    
  &=&  \frac{\alpha_s}{2\pi} \left ( -\frac{2}{\epsilon_c} + \ln \frac{e^\gamma \mu^2}{4\pi\mu_c^2 }\right )
       \biggr ( {\mathcal F}_q \otimes T_F + {\mathcal F }_{\Delta q} \otimes T_\Delta + {\mathcal  F }_{g} \otimes T_{G+}    \biggr ) (x),  
\nonumber\\
  \Delta \bar q(x)  &=&  \frac{\alpha_s }{2\pi} \left ( -\frac{2}{\epsilon_c} + \ln \frac{e^\gamma \mu^2}{4\pi\mu_c^2 }\right ) 
     \int \frac{d\xi}{\xi} \biggr \{  P_{qq}(z  ) \bar q (\xi)
       +  \frac{1}{2} \biggr [ z^2+ (1-z )^2\biggr ]   G(\xi) \biggr \}  
\nonumber\\
    &=&  \frac{\alpha_s }{2\pi} \left ( -\frac{2}{\epsilon_c} + \ln \frac{e^\gamma \mu^2}{4\pi\mu_c^2 }\right ) 
      \biggr ( P_{qq}\otimes \bar q + P_{qg} \otimes G \biggr ) (x),         
\label{SUBTFQ}     
\end{eqnarray}
with $z=x/\xi$. Here we define five convolutions ${\mathcal F}$'s for short notations. 
The derivative of $\Delta T_F(x,x)$ with $\mu$ gives the evolution kernel of $T_F(x,x)$ derived 
in \cite{BMP, MW, SchZh, KangQiu2}. Adding the contribution for the subtraction in Eq.(\ref{SUBCE}) to the one-loop 
correction, we find that all divergent contributions with the single-pole in $\epsilon$ 
are cancelled. Hence, the final results are finite. 
\par 
Before ending this section, it should be mentioned that only the contributions from Fig.2, Fig.5 and Fig.6 to the 
differential cross-section weighted with ${\mathcal O}_1$  
have been studied in \cite{VoYu}, where the integration over $x$ has been performed partly. Comparing with ours 
the results in \cite{VoYu} are incomplete for the chirality-even contributions.

\vskip20pt
\noindent 
{\bf 4.  The One-Loop Correction II }

In this section, we consider the real corrections involving twist-3 chirality-odd operators.  There are hard-pole- and soft-pole contributions. There is no contributions involving the twist-3 purely gluonic matrix elements and twist-2 gluon distribution functions.   The contributions at one-loop 
are from diagrams which have the same patten as given Fig.1, where the roles of $h_A$ and $h_B$ are exchanged and the direction of quark lines are reversed. Keeping this in mind, the hard-pole contributions are from Fig.\ref{HP1} and Fig.\ref{HP3}. The soft-gluon contributions are from Fig.\ref{SP1} and the soft-quark-pole contributions are from the diagrams in the first row of Fig.\ref{SFP1}. The calculations 
are similar to those in the last section. In the below we will only list our results from these diagrams without giving 
the details about the calculations. In Subsect. 4.1. we give the results from the mentioned diagrams and the virtual corrections. In Subsect.4.2. we study the subtraction. 

\par\vskip10pt
\noindent 
{\bf 4.1. The Unsubtracted Contributions}  
\par 
The hard-pole contributions from Fig.\ref{HP1} are:  
 \begin{eqnarray} 
  \frac{ d\sigma \langle {\mathcal O}_1 \rangle }{ d x d Q^2 }\biggr\vert_{Fig.\ref{HP1}}    
   &=& \frac{\vert s_\perp \vert^2 A_{\Gamma 1}} {4 x s Q^2}   \frac{\alpha_s     }{4 \pi N_c^2  }     
     F_D   \int \frac{d\xi_1 d\xi_2}{\xi_1\xi_2}   h_1(x_a) T_F^{(\sigma)}  (y_b,y_0)   \biggr (  N_c^2 \delta (1-\xi_1)\delta (1-\xi_2) \frac{2}{\epsilon}
\nonumber\\          
     &&    + {\mathcal A}_{1\sigma }(\xi_1,\xi_2) \biggr )  , 
\nonumber\\
    \frac{ d\sigma \langle {\mathcal O}_2 \rangle }{ d x d Q^2 }\biggr\vert_{Fig.\ref{HP1}}    
   &=& \frac{  \vert s_\perp \vert^2 A_{\Gamma 2}} {4 x s } \frac{\alpha_s    }{4 \pi N_c^2  }     
     F_D  \int \frac{d\xi_1 d\xi_2}{\xi_1\xi_2}   h_1(x_a) T_F^{(\sigma)}  (y_b,y_0 )
     \biggr \{   2  N_c^2 \delta (1-\xi_1)\delta (1-\xi_2) \left (\frac{2}{\epsilon}\right )^2 
\nonumber\\          
     &&   -2  N_c^2  \frac{2}{\epsilon} \frac{\delta(1-\xi_1) }{(1-\xi_2)_+} 
        - 2 \xi_1 (N_c^2+\xi_1-1)    \frac{2}{\epsilon} \frac{\delta (1-\xi_2)}{(1-\xi_1)_+} + {\mathcal A}_{2\sigma}(\xi_1,\xi_2) \biggr \},
\label{HP1o}                    
\end{eqnarray}
with $y_0  = \xi_2 y_b$. 
We note that in the second equation in Eq.(\ref{HP1o}) there is a term with the double pole in $\epsilon$, while the first 
equation contains  only a single-pole in $\epsilon$.

\par
The hard-pole contributions from Fig.\ref{HP3} are: 
\begin{eqnarray} 
   \frac{ d\sigma \langle {\mathcal O}_1 \rangle }{ d x d Q^2 }\biggr\vert_{Fig.\ref{HP3}}   
   &=& \frac{ \vert s_\perp \vert^2 A_{\Gamma 1}} {4 x s Q^2  }   \frac{\alpha_s     }{4 \pi N_c^2  }     
    F_D    \int \frac{d\xi_1 d\xi_2}{\xi_1\xi_2}    h_1(x_a) \biggr (  T_F^{(\sigma)} (y_b-y_0, -y_0) 
     {\mathcal B}_{1\sigma 1} (\xi_1,\xi_2) 
\nonumber\\     
   &&  + T_F^{(\sigma)} (y_0,y_0- y_b)  {\mathcal B}_{1\sigma 2} (\xi_1,\xi_2)  \biggr ), 
\nonumber\\
   \frac{ d\sigma \langle {\mathcal O}_2 \rangle }{ d x d Q^2 }\biggr\vert_{Fig.\ref{HP3}}   
   &=& \frac{ \vert s_\perp \vert^2 A_{\Gamma 2}} {4 x s }   \frac{\alpha_s     }{4 \pi N_c^2  }     
    F_D    \int \frac{d\xi_1 d\xi_2}{\xi_1\xi_2}   h_1(x_a) \biggr \{  
   -2 \delta (1-\xi_1) T_F^{(\sigma)} (y_0,y_0- y_b)  \frac{2}{\epsilon}(1-\xi_2)   
\nonumber\\        
      &&   + T_F^{(\sigma)} (y_b-y_0, -y_0) 
     {\mathcal B}_{2\sigma 1} (\xi_1,\xi_2) 
   + T_F^{(\sigma)} (y_0,y_0- y_b)  {\mathcal B}_{2\sigma 2} (\xi_1,\xi_2)  \biggr \}.  
\label{HP2o}                 
\end{eqnarray}  
In Eq.(\ref{HP2o}) the first equation does not contain a pole in $\epsilon$, while the second 
equation contains  only a single-pole in $\epsilon$.

\par
The soft-gluon-pole contributions from Fig.\ref{SP1} are: 
\begin{eqnarray} 
  \frac{ d\sigma \langle {\mathcal O}_1 \rangle }{ d x d Q^2 }\biggr\vert_{Fig.\ref{SP1} }   
   &=& \frac{\vert s_\perp \vert^2 A_{\Gamma 1}} {4 x s Q^2}  \frac{\alpha_s     }{4 \pi N_c^2  }     
     F_D    \int \frac{d\xi_1 d\xi_2}{\xi_1\xi_2}   h_1(x_a)   T_F^{(\sigma)} (y_b,y_b) 
  \biggr ( - \delta (1-\xi_1) \delta (1-\xi_2) \frac{2}{\epsilon}  
\nonumber\\
    &&   +{\mathcal C}_{1\sigma} (\xi_1,\xi_2) \biggr ), 
\nonumber\\
   \frac{ d\sigma \langle {\mathcal O}_2 \rangle }{ d x d Q^2 }\biggr\vert_{Fig.\ref{SP1} }    
   &=& \frac{\vert s_\perp \vert^2 A_{\Gamma 2}} {4 x s }  \frac{\alpha_s     }{4 \pi N_c^2  }     
     F_D    \int \frac{d\xi_1 d\xi_2}{\xi_1\xi_2}   h_1(x_a)  T_F^{(\sigma)} (y_b, y_b) 
  \biggr [ - 2 \delta (1-\xi_1)\delta (1-\xi_2) \biggr ( \left (\frac{2}{\epsilon} \right )^2 
\nonumber\\       
      && - \frac{2}{\epsilon} \biggr )  +  2 \xi_2 \delta (1-\xi_1) \frac{2}{\epsilon}  \frac{1}{(1-\xi_2)_+}  
    + 2\xi_1^2 \delta (1-\xi_2)  \frac{2}{\epsilon} \frac{1}{(1-\xi_1)_+} + {\mathcal C}_{2\sigma} (\xi_1,\xi_2)  \biggr ].   
\label{SGPo}                                          
\end{eqnarray}
In the second equation of Eq.(\ref{SGPo}) there is a term with the double-pole in $\epsilon$.

\par
The soft-quark-pole contributions from the first row of Fig.\ref{SFP1} are: 
\begin{eqnarray} 
   \frac{ d\sigma \langle {\mathcal O}_i \rangle }{ d x d Q^2 }\biggr\vert_{Fig.7 }    
   &=& \frac{\vert s_\perp \vert^2 A_{\Gamma i}} {4 x s Q^{2(i-2)} }   \frac{\alpha_s     }{4 \pi N_c^2  }     
       \int \frac{d\xi_1 d\xi_2}{\xi_1\xi_2}   h_1(x_a)   T_F^{(\sigma)} (0,-y_b) 
      {\mathcal D}_{i\sigma} (\xi_1,\xi_2),
\label{SFPo}                           
\end{eqnarray}  
for $i=1,2$. 
These contributions are finite. The complete real corrections are the sum of the results given by Eq.(\ref{HP1o}), Eq.(\ref{HP2o}), Eq.(\ref{SGPo}) and Eq.(\ref{SFPo}). 
\par

As discussed, the virtual corrections are obtained by the replacement specified with Eq.(\ref{VREP}). Therefore, the virtual 
corrections for the chirality-odd contributions are: 
\begin{eqnarray} 
  \frac{ d\sigma \langle {\mathcal O}_1 \rangle }{ d x d Q^2 }\biggr\vert_{vir.}  
   &=& -\frac{  \vert s_\perp \vert^2 A_{\Gamma 1}} {4 x s Q^2}   \frac{\alpha_s C_F  }{2 \pi N_c  }     
     F_D   \int \frac{d\xi_1 d\xi_2}{\xi_1\xi_2}  \delta(1-\xi_1) \delta (1-\xi_2)  h_1(x_a)   T_F^{(\sigma)} (y_b,y_b)  
  \biggr ( \frac{2}{\epsilon} + \frac{3}{2} \biggr ), 
\nonumber\\
   \frac{ d\sigma \langle {\mathcal O}_2 \rangle }{ d x d Q^2 }\biggr\vert_{vir.}    
   &=& - \frac{ \vert s_\perp \vert^2 A_{\Gamma 2}} {4 x s }   \frac{\alpha_s  C_F  }{2 \pi N_c  }  F_D    
     \int \frac{d\xi_1 d\xi_2}{\xi_1\xi_2}  \delta(1-\xi_1) \delta (1-\xi_2) h_1(x_a)  T_F^{(\sigma)} (y_b,y_b) 
\nonumber\\     
   && \cdot \biggr ( 2 \left ( \frac{2}{\epsilon}\right )^2 +\frac{2}{\epsilon} +5 -\pi^2 \biggr ) .                         
\end{eqnarray}
Since the tree-level chirality-odd contribution to the differential cross-section weighted with ${\mathcal O}_1$ is proportional to $\epsilon$ in Eq.(\ref{TreeC}), the corresponding one-loop virtual correction has only single-pole 
in $\epsilon$.

\par\vskip10pt
\noindent 
{\bf 4.2. The Subtraction for the Chirality-Odd Contributions}
\par 

Summing the various contributions, we obtain the divergent part of the one-loop corrections to the chirality-odd contributions:  
\begin{eqnarray} 
  \frac{ d\sigma ({\mathcal O}_1 ) }{ d x d Q^2 } \biggr\vert_{div.}    
   &=&  \frac{2}{\epsilon} \times 0, 
\nonumber\\
 \frac{ d\sigma ({\mathcal O}_2 ) }{ d x d Q^2 }\biggr\vert_{div.}   
   &=& \frac{\vert s_\perp \vert^2    A_{\Gamma 2}} {4 x s }  \frac{\alpha_s     }{4 \pi N_c^2  }     
     F_D    \frac {2}{\epsilon} \int \frac{d\xi_1 d\xi_2}{\xi_1\xi_2}  \biggr \{  h_1(x_a)  T_F^{(\sigma)} (y_b,y_b) 
\nonumber\\          
     &&  \delta (1-\xi_1)\delta(1-\xi_2)  ( 3 - N_c^2) + \delta (1-\xi_1) h_1(x_a)   \biggr [ -2  \frac{1}{(1-\xi_2)_+}
\nonumber\\   
     &&   \biggr ( N_c^2 T_F^{(\sigma)} (y_b,y)  -\xi_2  T_F^{(\sigma)} (y_b,y_b) \biggr )  -2 (1-\xi_2)     T_F^{(\sigma)} (y, y-y_b) \biggr ] 
\nonumber\\
    && - \frac{2}{\epsilon} \delta (1-\xi_2) T_F^{(\sigma)}(y_b,y_b) h_1(x_a)   (N_c^2-1) \frac{2 \xi_1}{(1-\xi_1)_+} \biggr\}.
\label{DCO}                               
\end{eqnarray}
We notice that there is no divergence in the chirality-odd contribution to the differential cross-section weighted with 
${\mathcal O}_1$ in the sum. 
In the chirality-odd contribution to the differential cross-section weighted with 
${\mathcal O}_2$ the double-pole terms in $\epsilon$ are cancelled, the remaining divergence is with the single-pole in 
$\epsilon$.  

\par 
Similar to the case of the chirality-even contributions, the divergence in the sum is from the momentum region where the parton in the intermediate states
or the exchanged gluon in the virtual correction  
is collinear to $h_A$ or $h_B$.  These collinear contributions are already included in the hadronic matrix elements 
in the tree-level results. Therefore, a subtraction is needed to avoid the double-counting.
\par 
The subtraction procedure is the same as discussed for the chirality-even contribution. We make the replacement in our 
tree-level results:
\begin{equation} 
   T_F^{(\sigma)} (x,x) \to  T_F^{(\sigma)} (x,x) -\Delta  T_F^{(\sigma)} (x,x), \quad 
     h_1(x) \to h_1 (x) -\Delta h_1(x) 
\end{equation} 
and obtain the contributions of the subtraction: 
 \begin{eqnarray} 
  \Delta  \frac{  d\sigma ({\mathcal O}_1 ) }{ dx  d Q^2 } 
   &=& \frac{A_{\Gamma 1} \vert s_\perp \vert^2} {8 x s Q^2 N_c} \biggr (-\frac{\epsilon}{2} \biggr )    \int \frac{d\xi_1 d\xi_2}{\xi_1\xi_2} \delta (1-\xi_1) 
      \delta (1-\xi_2)  \biggr ( \Delta h_1(x_a) T_F^{(\sigma)} (y_b,y_b) 
\nonumber\\      
     &&  + h_1(x) \Delta T_F^{(\sigma)} (y_b,y_b) \biggr ) ,                                   
\nonumber\\
  \Delta \frac{   d\sigma ({\mathcal O}_2 )} {  d x  d Q^2 } 
   &=&   - \frac{A_{\Gamma 2}\vert s_\perp \vert^2 (2-\epsilon) } {8 x s N_c}  \int\frac{d\xi_1 d\xi_2}{\xi_1\xi_2} \delta (1-\xi_1) 
      \delta (1-\xi_2)     
    \biggr ( \Delta h_1(x_a) T_F^{(\sigma)} (y_b,y_b) 
\nonumber\\      
     &&  + h_1(x) \Delta T_F^{(\sigma)} (y_b,y_b) \biggr ). 
\label{SUBCO} 
\end{eqnarray}
These contributions should be added to the one-loop corrections in the previous subsection.  We notice that the collinear 
contributions are not always divergent. An example is the case of the chirality-odd contribution given by the first equation 
in Eq.(\ref{DCO}). With the correct factorization this corresponds to the fact that the contribution of the subtraction in the first equation 
of Eq.(\ref{SUBCO}) is finite at one-loop.

\par 
Again, $\Delta  T_F^{(\sigma)} (x,x)$ and $\Delta h_1(x)$ are determined by their evolution, respectively. 
The evolution of $T_F^{(\sigma)}$ has been studied in \cite{KangQiu2,Beli1,MWZh}. From the evolution 
we have:    
\begin{eqnarray} 
\Delta  T_F^{(\sigma)} (x,x,\mu)    
     & =& \frac{\alpha_s}{2\pi} \left ( -\frac{2}{\epsilon_c} + \ln \frac{e^\gamma \mu^2}{4\pi\mu_c^2 }\right )  \biggr \{ -\frac{N_c^2 +3}{4 N_c}T_F^{(\sigma)}  (x,x,\mu) 
         +  \int_x^1 \frac{dz}{z} \frac{1}{(1-z)_+} \biggr  ( N_c T_F^{(\sigma)} (x,\xi) 
\nonumber\\
        &&         
         -\frac{z}{N_c} T_F^{(\sigma)} (\xi,\xi)\biggr  ) 
      + \frac{1}{N_c} \int_0 ^{1-x} d\xi \frac{\xi}{(\xi+x)^2} T_F^{(\sigma)} (x,-\xi) \biggr \}
\nonumber\\      
     & =& \frac{\alpha_s}{2\pi} \left ( -\frac{2}{\epsilon_c} + \ln \frac{e^\gamma \mu^2}{4\pi\mu_c^2 }\right ) 
         \biggr ( {\mathcal F}_{\sigma}\otimes T_F^{(\sigma)} \biggr )  (x) , 
\label{SG}      
\end{eqnarray}       
with $z =x/\xi$. Taking the derivative of $\Delta  T_F^{(\sigma)} (x,x,\mu)$ we obtain the evolution of 
$T_F^{(\sigma)} (x,x,\mu)$. The evolution of $h_1$ has been determined in \cite{WVTE}. From the result there we have: 
\begin{eqnarray} 
   \Delta h_1 (x)  &=& \frac{\alpha_s }{2\pi} \left ( -\frac{2}{\epsilon_c} + \ln \frac{e^\gamma \mu^2}{4\pi\mu_c^2 }\right ) 
     C_F \int \frac{d\xi}{\xi}   \biggr ( \frac{2 z }{(1- z)_+} + \frac{3}{2} \delta (1- z)  \biggr )  h_1  (\xi) 
\nonumber\\
    &=& \frac{\alpha_s }{2\pi} \left ( -\frac{2}{\epsilon_c} + \ln \frac{e^\gamma \mu^2}{4\pi\mu_c^2 }\right ) 
        \biggr ( P_{\perp q} \otimes h_1 \biggr )  (x).     
\end{eqnarray}    
As in Subsection 3.4. we define here two convolutions for short notations. 

\par
With the given $ \Delta  T_F^{(\sigma)}$ and $\Delta h_1$ in the above one can perform the subtraction 
with Eq.(\ref{SUBCO}). For the differential cross-section weighted with ${\mathcal O}_2$, 
we realize that all divergent parts with the pole in $\epsilon$ 
are exactly cancelled after the subtraction. For the differential cross-section weighted with ${\mathcal O}_1$, although there is no collinear divergence, the subtraction 
is finite here.

\par\vskip20pt
\noindent 
{\bf 5. The Final Results} 
\par
To sum our results in previous sections,  we introduce two functions as the sums of evolutions combined 
with other distributions:  
\begin{eqnarray} 
   {\mathcal A}(x_a,y_b ) &=&  \bar q(y_b)           
          \biggr ( {\mathcal F}_q \otimes T_F + {\mathcal F }_{\Delta q} \otimes T_\Delta 
         + {\mathcal  F }_{g} \otimes T_{G+}   \biggr ) (x_a )  + T_F (x_a,x_a)\biggr ( P_{qq}\otimes \bar q + P_{qg} \otimes G \biggr ) (y_b ), 
\nonumber\\
     {\mathcal B}(x_a,y_b ) &=&  h_1(x_a) \biggr ( {\mathcal F}_{\sigma} \otimes T_F^{\sigma} \biggr ) (y_b,y_b) 
 + T_F^{(\sigma)} ( y_b, y_b) \biggr ( P_{\perp q} \otimes h_1 \biggr )( x_a) .      
\end{eqnarray}
The various evolutions can be found in the subsection 3.4 and 4.2.       
 The constants $A_{\Gamma 1}$ and $A_{\Gamma 2}$ with $d=4$ are:
\begin{equation} 
   A_{\Gamma 1} = \frac{1}{6 \pi}, \quad A_{\Gamma 2} = \frac{1}{240\pi}. 
\end{equation}    
Our final result for the differential cross-section weighted with ${\mathcal O}_1$ is: 
\begin{eqnarray} 
   \frac {d \sigma\langle {\mathcal O}_1 \rangle }{ dx d Q^2 } &=&  -\frac{\vert s_\perp\vert^2 A_{\Gamma 1} }{4 x s Q^2 N_c} 
      \int \frac{d\xi_1 d\xi_2}  {\xi_1 \xi_2}  \biggr \{ \delta (1-\xi_1) \delta (1-\xi_2 ) \biggr [  \bar q (y_b) T_F (x_a,x_a) 
         -\frac{\alpha_s}{2\pi}   {\mathcal A}(x_a,y_b) \ln \frac{ e \mu^2}{Q^2} 
         \nonumber\\    
    && + \frac{\alpha_s C_F }{2\pi} \biggr (  \pi^2 -5 \biggr ) \bar q (y_b) T_F (x_a,x_a)  
  +\frac{\alpha_s} {4\pi} \biggr ( -  {\mathcal B} (x_a,y_b)    + 3 C_F h_1 (x_a) T_F^{(\sigma)} (y_b,y_b)   \biggr ) \biggr ] 
\nonumber\\   
     && +\frac{\alpha_s}{8\pi} \frac{1}{x_a}  \delta (1-\xi_2)  (2\xi_1^2-2\xi_1+1)   \biggr ( 3  T_{G+} (x_a,x_a) - 2 T_{G-}(x_a,0) \biggr )   \biggr \} + 
       \frac {d \sigma\langle {\mathcal O}_1 \rangle }{ dx d Q^2 }
\biggr\vert_F.  
\end{eqnarray} 
The last term stands for the sum of all finite parts in previous sections.                
The final result for the differential cross-section weighted with ${\mathcal O}_2$ is: 
\begin{eqnarray} 
   \frac {d \sigma\langle {\mathcal O}_2 \rangle }{ dx d Q^2 } &=&  -\frac{\vert s_\perp\vert^2 A_{\Gamma 2} }{4 x s  N_c} 
      \int \frac{d\xi_1 d\xi_2}  {\xi_1 \xi_2}  \biggr \{ \delta (1-\xi_1) \delta (1-\xi_2 ) \biggr [  \bar q (y_b) T_F (x_a,x_a) 
         -\frac{\alpha_s}{2\pi}   {\mathcal A}(x_a,y_b) \ln \frac{\mu^2}{Q^2} 
\nonumber\\         
      && + \frac{\alpha_s C_F }{2\pi} \biggr (  \pi^2  -8  \biggr ) \bar q (y_b) T_F (x_a,x_a)-h_1 (x_a) T_F^{ (\sigma)} (y_b,y_b) + \frac{\alpha_s} {2\pi}
      {\mathcal B} (x_a,y_b) \ln \frac{e \mu^2}{Q^2}            
\nonumber\\    
     &&     + \frac{\alpha_s C_F }{2\pi} \biggr ( 5 - \pi^2 \biggr )  h_1(x_a) T_F^{(\sigma)} (y_b,y_b)  \biggr ]   +\frac{\alpha_s}{8\pi} \frac{1}{x_a}  \delta (1-\xi_2)  (2\xi_1^2-2\xi_1+1)  
\nonumber\\ 
   &&  \biggr ( 3  T_{G+} (x_a,x_a) - 2 T_{G-}(x_a,0) \biggr )   \biggr \} + 
       \frac {d \sigma\langle {\mathcal O}_2 \rangle }{ dx d Q^2 }
\biggr\vert_F. 
\end{eqnarray}                
In these results the contribution with the combination $( 3 T_{G+}- 2T_{G-})$ is the extra contribution discussed after Eq.(\ref{TGPM}). 
The final results are finite. 
\par 
The finite parts in the above are given by: 
\begin{eqnarray}
    \frac{ d\sigma \langle {\mathcal O}_i\rangle  }{ d x d Q^2}\biggr\vert_F &=& \frac{ \alpha_s \vert s_\perp \vert^2 A_{\Gamma i}}{16\pi  x s (Q^2)^{2-i}  N_c^2}  \int \frac{ d\xi_1 d\xi_2}{ \xi_1\xi_2} \biggr \{  \bar q(y_b)  \biggr [ T_F (x_a,z_a) 
    {\mathcal A}_{iF} (\xi_1,\xi_2) + T_{\Delta} (x_a,z_a) {\mathcal A }_{i\Delta }(\xi_1,\xi_2)
\nonumber\\
   &&+T_F(-x_\xi, z_a) {\mathcal C}_{iF1} (\xi_1,\xi_2) + T_F(-z_a, x_\xi) {\mathcal C}_{iF2} (\xi_1,\xi_2)
   + T_\Delta (-x_\xi, z_a) {\mathcal C}_{i\Delta 1} (\xi_1,\xi_2) 
\nonumber\\   
  &&  + T_\Delta (-z_a, x_\xi) {\mathcal C}_{i\Delta 2} (\xi_1,\xi_2) +T_F (x_a,x_a) {\mathcal D}_i (\xi_1,\xi_2) 
 + \frac{1}{2 x_a}  \biggr ( \frac{3}{2} T_{G+}(x_a,x_a)-T_{G-}(x_a,0) \biggr ) 
\nonumber\\ 
 && {\mathcal G}_{i+} (\xi_1,\xi_2) 
    + \frac{1}{2 x_a}  \biggr (T_{G-}(x_a, 0)- \frac{1}{2} T_{G+}(x_a,x_a)  \biggr ) 
       {\mathcal G}_{i-} (\xi_1,\xi_2) \biggr ]           
\nonumber\\
  && + G(y_b) \biggr [  T_F (x_a,z_a) 
    {\mathcal B}_{iF} (\xi_1,\xi_2) +  T_{\Delta} (x_a,z_a) {\mathcal B }_{i\Delta}(\xi_1,\xi_2) + T_F (x_a,x_a) {\mathcal E}_i (\xi_1,\xi_2)\biggr ]  
\nonumber\\
  && + \biggr ( \bar q (y_b) +\frac{1}{2 C_F} G(y_b) \biggr ) \biggr [ T_F (-x_a, 0) {\mathcal F}_{iF} (\xi_1,\xi_2)  
       + T_{\Delta} (-x_a,0) {\mathcal F}_{i\Delta} (\xi_1,\xi_2) \biggr ] 
\nonumber\\ 
   && +  h_1(x_a) \biggr [  T_F^{(\sigma)}  (y_b,y_0) {\mathcal A}_{i\sigma} (\xi_1,\xi_2) +  T_F^{(\sigma)} (y_b-y_0, -y_0) 
     {\mathcal B}_{i\sigma 1} (\xi_1,\xi_2) 
\nonumber\\     
   &&  + T_F^{(\sigma)} (y_0,y_0- y_b)  {\mathcal B}_{i\sigma 2} (\xi_1,\xi_2)  + T_F^{(\sigma)} (y_b,y_b) {\mathcal C}_{i\sigma} 
   (\xi_1,\xi_2)   
 +   T_F^{(\sigma)} (0,-y_b) 
      {\mathcal D}_{i\sigma} (\xi_1,\xi_2)
   \biggr ]    
    \biggr\}, 
\label{FIPA}                 
\end{eqnarray}
with 
\begin{equation} 
  z_a = \frac{x \xi_2}{1-\xi_1 (1-\xi_2) }, \quad x_\xi =x_a-z_a, \quad y_0 = \xi_2 y_b. 
\end{equation}  
The functions ${\mathcal A}$'s to ${\mathcal G}$'s are given in Appendix.  

\par 

As discussed in previous sections about subtractions, the evolutions of $T_F(x,x)$, $T_F^{(\sigma)}(x,x)$ and $h_1 (x)$ 
in our short notations 
are given by: 
\begin{eqnarray}
\mu \frac{\partial  T_F(x,x,\mu) }{\partial \mu} &=&   \frac{\alpha_s}{\pi} 
       \biggr ( {\mathcal F}_q \otimes T_F + {\mathcal F }_{\Delta q} \otimes T_\Delta + {\mathcal  F }_{g} \otimes T_{G+}    \biggr ) (x),  
\nonumber\\ 
 \mu \frac{\partial   T_F^{(\sigma)} (x,x,\mu) }{\partial \mu}    
     & =&  \frac{\alpha_s}{\pi} 
         \biggr ( {\mathcal F}_{\sigma}\otimes T_F^{(\sigma)} \biggr )  (x) , 
\nonumber\\ 
   \mu\frac{\partial  h_1 (x,\mu)}{\partial \mu}   &=& \frac{\alpha_s }{\pi} 
        \biggr ( P_{\perp q} \otimes h_1 \biggr )  (x).     
\end{eqnarray}    
The evolution of $T_F(x,x)$ is given in \cite{BMP, MW, SchZh, KangQiu2}. The evolution of $T_F^{(\sigma)}$ 
and $h_1$ are derived in \cite{KangQiu2,MWZh} and \cite{WVTE}, respectively. With these evolutions and those of the standard parton 
distribution functions one can easily verify that our final results do not depend on the renormalization scale $\mu$. 
\par 

\par\vskip20pt
\noindent 
{\bf 6. Summary}

\par 
We have performed one-loop calculations for the two weighted differential cross-sections. They are transverse-spin dependent. 
These differential cross-sections are factorized with hadronic matrix elements defined not only with twist-2 operators but also with twist-3 operators. In our results all collinear contributions, which can be divergent, are 
factorized into hadronic matrix-elements. The final results are finite. Our work gives an example of twist-3 factorization at one-loop, in particular, of the factorization with chirality-odd twist-3 operators in the first time. With our results 
SSA's can be predicted more precisely than with tree-level results, or one can extract from experiment, e.g. at RHIC, 
twist-3 parton distributions more precisely by measuring the two observables studied here. These distributions will help to understand the inner structure of hadrons. 
Besides twist-3 parton distributions, one can also use our results to extract the twist-2 transversity distribution, which 
is still not well-known.            
\par 
In this work, the two weighted differential cross-sections for SSA are constructed in the way that their virtual corrections, as discussed in Introduction, are determined by the corrections of the quark form factor. It is noted that one can construct 
more observables than the two here. The virtual corrections of these observables may not be determined by the quark form factor and 
can be complicated. 
We leave the study of one-loop corrections for such observables in the future.

\par\vskip20pt
\noindent
{\bf Acknowledgments}
\par
The work is supported by National Nature
Science Foundation of P.R. China(No.11275244, 11675241, 11605195). The partial support from the CAS center for excellence in particle 
physics(CCEPP) is acknowledged.

\par\vskip20pt
\renewcommand{\theequation}{A.\arabic{equation}}
\setcounter{equation}{0}

\noindent
{\bf Appendix:}
\par 
We give here all functions appearing in the finite part of our results in Eq.(\ref{FIPA}). These functions are:  
\begin{eqnarray} 
  {\mathcal A}_{1F} (\xi_1,\xi_2) &=&   \delta (1-\xi_1) 
      (N_c^2 +\xi_2-1) 
 \biggr ( \xi_2  - 1-  (1+\xi_2^2) L_1 (\xi_2)   \biggr ) - \delta (1-\xi_2) N_c^2 (1+\xi_1)  L_2 (\xi_1) 
\nonumber\\
    && 
       -  (N_c^2 + \tilde \xi_2 -1 ) \cdot \frac{ \xi_1 + \tilde \xi_2^3 +\tilde\xi_2 -1 }{ \tilde \xi_2 (\xi_1-1)_+ (\xi_2-1)_+},
\nonumber\\
  {\mathcal A}_{1\Delta } (\xi_1,\xi_2) &=& \delta (1-\xi_2) N_c^2  
   (1-\xi_1) L_2 (\xi_1) -   (N_c^2 + \tilde \xi_2 -1 ) \frac{ \xi_1 -\tilde \xi_2^3 +\tilde\xi_2 -1 }{ \tilde \xi_2 (\xi_1-1)_+ (\xi_2-1)_+} , 
\nonumber\\
     {\mathcal A}_{2F} (\xi_1,\xi_2) &=&   -2 N_c^2    \delta (1-\xi_1)\delta(1-\xi_2)  
 + \delta (1-\xi_1) 
      \frac{ N_c^2 +\xi_2-1}{(1-\xi_2)_+ }  \biggr ( (1+\xi_2^2)(\xi_2-1) L_1(\xi_2) 
           + 2\xi_2  \biggr ) 
\nonumber\\
    &&           
+ \delta (1-\xi_2) N_c^2 \frac{ 1+\xi_1}{(1-\xi_1)_+} 
  \biggr ( (\xi_1-1)  L_2 (\xi_1) +1  \biggr ) 
       -  (N_c^2 + \tilde \xi_2 -1 ) \frac{ \xi_1 +\tilde\xi_2^2 }{ (\xi_1-1)_+ (\xi_2-1)_+},   
\nonumber\\     
    {\mathcal A}_{2\Delta } (\xi_1,\xi_2) &=& \delta (1-\xi_2) N_c^2  \biggr (  (1-\xi_1) L_2 (\xi_1) - 1 \biggr )  -  (N_c^2 + \tilde \xi_2 -1 ) \cdot \frac{ \xi_1 -\tilde\xi_2^2 }{ (\xi_1-1)_+ (\xi_2-1)_+},
\nonumber\\     
     {\mathcal B}_{1F} (\xi_1,\xi_2) &=& \frac{1}{2C_F} \biggr [ \delta (1-\xi_1) 
  (N_c^2 (1-\xi_2) -1)\biggr (  -L_1(\xi_2)(1-\xi_2)   ( 2\xi_2^2 -2 \xi_2 +1) + 2 \xi_2 (\xi_2-1) \biggr )     
\nonumber\\   
    &&  +\frac{ (1-\txi2 )N_c^2 -1}{\txi2 (1-\xi_1)_+} \xi_1 (2\xi_1\txi2 -\xi_1\txi2 ^2 -\xi_1 -2 \txi2 ^3 + 3 \txi2 ^2 -3\txi2 +1) \biggr ],  
\nonumber\\
  {\mathcal B}_{1\Delta } (\xi_1,\xi_2) &=& \frac{1}{2C_F} \frac{ (1-\txi2 )N_c^2 -1}{\txi2 (1-\xi_1)_+} \xi_1 (2\xi_1\txi2 -\xi_1\txi2 ^2 -\xi_1 + 3 \txi2 ^2 -3\txi2 +1), 
\nonumber\\     
     {\mathcal B}_{2F} (\xi_1,\xi_2) &=& \frac{1}{2C_F} \biggr [ \delta (1-\xi_1) (N_c^2 (1-\xi_2) -1)  
 \biggr ( - (1-\xi_2)    ( 2\xi_2^2-2\xi_2 +1)L_1(\xi_2) + (2 \xi_2-1)^2  \biggr ) 
\nonumber\\      
    && - (N_c^2 (1-\txi2 ) -1) \frac{\xi_1 ( 2\txi2 ^2 -2 \txi2 +1)}{ (1-\xi_1)_+ } \biggr ], 
\nonumber\\
  {\mathcal B}_{2\Delta } (\xi_1,\xi_2) &=& \frac{1}{2C_F}(N_c^2 (1-\txi2 ) -1) \frac{\xi_1 (2 \txi2 - 1)}{(1-\xi_1)_+ } ,   
\nonumber\\
    {\mathcal C}_{1F1} (\xi_1,\xi_2) &=&  (2\xi_1-1) \delta (1-\xi_2) L_2(\xi_1) (1-\xi_1)  
     +\frac{ 2\xi_1 +\txi2 -2}{\txi2 (1-\xi_2)_+}
     - N_c\frac{\xi_1}{\txi2 ^2} (2\xi_1 +\txi2 -2) ( \txi2 ^2 -2\txi2 +2),     
\nonumber\\
  {\mathcal C}_{1\Delta1 } (\xi_1,\xi_2) &=&  \delta (1-\xi_2)  L_2(\xi_1) (1-\xi_1)   +\frac{1}{(1-\xi_2)_+ }- N_c \frac{\xi_1}{\txi2 } (\txi2 ^2- 2\txi2 + 2),
\nonumber\\
   {\mathcal C}_{1F2} (\xi_1,\xi_2) &=& \frac{\xi_1 (1-\txi2 )^2 ( 2\xi_1 + \txi2 -2)}{\txi2 } + N_c\frac{\xi_1}{\txi2 ^2} (2\xi_1 +\txi2 -2) ( \txi2 ^2 -2\txi2 +2) 
\nonumber\\
  {\mathcal C}_{1\Delta 2 } (\xi_1,\xi_2) &=&  - \xi_1 (1- \txi2 )^2  
    - N_c \frac{\xi_1}{\txi2 } (\txi2 ^2- 2\txi2 + 2),
\nonumber\\
    {\mathcal C}_{2F1} (\xi_1,\xi_2) &=&  (2\xi_1-1) \delta (1-\xi_2)  
 \biggr (  L_2(\xi_1) (1-\xi_1)  -1 \biggr ) +\frac{ 2\xi_1 -2 \txi2 +1 }{ (1-\xi_2)_+} 
\nonumber\\ 
  && -N_c \frac{\xi_1}{\txi2 } ( -2 \xi_1 \txi2 + 4 \xi_1 +\txi2 ^2 -4 \txi2 +2)      
\nonumber\\
  {\mathcal C}_{2\Delta1 } (\xi_1,\xi_2) &=&  \delta (1-\xi_2) \biggr (  L_2(\xi_1) (1-\xi_1) -1 \biggr )  +\frac{2\txi2 -1 }{(1-\xi_2)_+ } + N_c \frac{\xi_1}{\txi2 } (-2 \xi_1 (\txi2 -1) -\txi2 ^2 ),   
\nonumber\\
   {\mathcal C}_{2F2} (\xi_1,\xi_2) &=& \xi_1 (1-\txi2 )  
    ( 2\xi_1 - \txi2 +1 ) +N_c\frac{\xi_1}{\txi2 } ( -2 \xi_1 \txi2 + 4 \xi_1 +\txi2 ^2 -4 \txi2 +2),  
\nonumber\\
  {\mathcal C}_{2\Delta 2 } (\xi_1,\xi_2) &=& \xi_1 (1- \txi2 )(2\xi_1 +\txi2 -1) + N_c \frac{\xi_1}{\txi2 } (-2 \xi_1 (\txi2 -1) -\txi2 ^2 ),
\nonumber\\
    {\mathcal D}_{1} (\xi_1,\xi_2) &=&   
         \delta (1-\xi_1) \biggr (  (1+\xi_2^2)\xi_2   L_1(\xi_2) 
   +\frac{1 +\xi_2-\xi_2^2 +\xi_2^3}{(1-\xi_2)_+} \biggr )  +\delta (1-\xi_2) \biggr ( (1+\xi_1^2)   L_2 (\xi_1) 
\nonumber\\
   &&         
           +\frac{1-\xi_1 +\xi_1^2 +\xi_1^3}{(1-\xi_1)_+} \biggr )
 + \frac{-\xi_1^2\txi2 + 2 \xi_1^2 + 2\xi_1 \txi2 ^2 -2 \xi_1 +\txi2 ^3 -2\txi2 ^2 +2 \txi2 }{(1-\xi_1)_+ (1-\xi_2)_+} , 
\nonumber\\
    {\mathcal D}_{2} (\xi_1,\xi_2) &=&  
          \delta (1-\xi_1) \biggr (  (1+\xi_2^2)  \xi_2 L_1(\xi_2)              
    + 1+\xi_2 \biggr ) +\delta (1-\xi_2) \biggr ( (1+\xi_1^2)  L_2 (\xi_1)  -\xi_1 (1+\xi_1) \biggr )
\nonumber\\
   && + \frac{ \xi_1^2 (2 \txi2 -1) +\xi_1 (2 \txi2 ^2 -3\txi2 +1) +\txi2 ^3 -2\txi2 ^2 +2 \txi2  }{(1-\xi_1)_+ (1-\xi_2)_+} , 
\nonumber\\
  {\mathcal E}_{1 } (\xi_1,\xi_2) &=& -\frac{N_c^2}{2 C_F}       \biggr [ \delta (1-\xi_1) \biggr (  ( 2\xi_2^2 -2 \xi_2 +1)
  \xi_2   (1-\xi_2) L_1(\xi_2)    -2\xi_2^3 +4 \xi_2^2 -2 \xi_2 +1 \biggr )
\nonumber\\   
   && +\frac{1}{(1-\xi_1)_+} ( \xi_1^4 +\xi_1^3 (3\txi2 -5) 
   +\xi_1^2 (6\txi2 ^2 -12 \txi2 +8) +\xi_1 ( 2\txi2 ^3 - 8\txi2 ^2 +10\txi2 -4) \biggr ],   
\nonumber\\
     {\mathcal E}_{2 } (\xi_1,\xi_2) &=&  -\frac{N_c^2}{2 C_F}    \biggr [ \delta (1-\xi_1)  \biggr ( \xi_2   (1-\xi_2)  ( 2\xi_2^2 -2 \xi_2 +1)L_1(\xi_2)    -4\xi_2^3 +6\xi_2^2 -3 \xi_2 +1  \biggr )
\nonumber\\
   && +\frac{1}{(1-\xi_1)_+} ( \xi_1^4 -2 \xi_1^3 \txi2 + \xi_1^2 ( 5\txi2 ^3 -2\txi2 -1) +\xi_1 \txi2 (2\txi2 ^2 -7 \txi2 +5  ) \biggr ],   
\nonumber\\
   {\mathcal F}_{1F} (\xi_1,\xi_2) &=& \xi_1  ( -2\xi_1^2  
    -3\xi_1 \txi2 +6\xi_1 -\txi2 ^2 +5\txi2 -5 ),  
 \nonumber\\
  {\mathcal F}_{1\Delta } (\xi_1,\xi_2) &=& \xi_1 ( \xi_1\txi2 -2\xi_1 
    +\txi2 ^2 -3 \txi2 +3),
\nonumber\\
   {\mathcal F}_{2F} (\xi_1,\xi_2) &=& \xi_1 ( -2\xi_1^2 +4 \xi_1 \txi2 -\xi_1 -\txi2 ^2 ) ,
\nonumber\\
  {\mathcal F}_{2\Delta } (\xi_1,\xi_2) &=& \xi_1 (\tilde \xi_2^2 -\xi_1), 
\nonumber\\
  {\mathcal G}_{1+} (\xi_1,\xi_2) &=& N_c  \biggr [  \delta (1-\xi_2) \biggr ( 2 (2\xi_1^2 -2 \xi_1 +1) (1-\xi_1) L_2(\xi_1)      + 4\xi_1^3 -4\xi_1^2 +2 \xi_1 +2  \biggr )
\nonumber\\
   && +\frac{1}{(1-\xi_2)_+} ( \xi_1^2 (8-4\txi2 ) +\xi_1 (8\txi2 -12) +2 \txi2 ^2 -8\txi2 -8 ) \biggr ] 
\nonumber\\
    {\mathcal G}_{1-} (\xi_1,\xi_2)  &=& N_c  \biggr [ \delta (1-\xi_2) \biggr ( 2 (2\xi_1^2 -2 \xi_1 +1) (1-\xi_1) L_2(\xi_1)      +  8\xi_1^3 -16\xi_1^2 +10 \xi_1  \biggr )
\nonumber\\
   && +\frac{1}{(1-\xi_2)_+} ( \xi_1^2 (16-12\txi2 ) + \xi_1 ( -8\txi2 ^2 + 32\txi2 -28) +10 \txi2 ^2 -24\txi2 +16 ) \biggr ]
\nonumber\\
  {\mathcal G}_{2+} (\xi_1,\xi_2) &=& N_c \biggr [  \delta (1-\xi_2) \biggr ( 2 (2\xi_1^2 -2 \xi_1 +1) (1-\xi_1) L_2(\xi_1)     + 4\xi_1^3 -8 \xi_1^2 +6 \xi_1  \biggr )
\nonumber\\
   && +\frac{1}{(1-\xi_2)_+} ( 4\xi_1^2 +\xi_1\txi2  ( 6\txi2  -10  ) -4\txi2 ^2 + 6\txi2 ) \biggr ] ,
\nonumber\\
    {\mathcal G}_{2-} (\xi_1,\xi_2)  &=& N_c  \biggr [ \delta (1-\xi_2) \biggr ( 2 (2\xi_1^2 -2 \xi_1 +1)  (1-\xi_1) L_2(\xi_1)      +  8\xi_1^3 -20\xi_1^2 +14 \xi_1 -2  \biggr )
\nonumber\\
   && +\frac{1}{(1-\xi_2)_+} ( \xi_1^2 (20-16\txi2 ) +\xi_1 (10 \txi2 ^2 -6 \txi2 -8) -8\txi2 ^2 + 10\txi2 ) \biggr ],
 \nonumber\\
   {\mathcal A}_{1\sigma} (\xi_1,\xi_2) &=&  N_c^2 \delta (1-\xi_1) \biggr ( \delta (1-\xi_2)  
   -  \frac{1}{(1-\xi_2)_+} \biggr )  -(N_c^2 +\xi_1-1) 
   \delta (1-\xi_2) \frac{\xi_1}{(1-\xi_1)_+}, 
\nonumber\\
  {\mathcal A}_{2\sigma} (\xi_1,\xi_2) &=& 2 \delta(1-\xi_1) N_c^2   (  L_1(\xi_2) + 1  ) 
 + 2 \xi_1 (N_c^2+\xi_1-1) \biggr (   \delta (1-\xi_2)L_2(\xi_1)
\nonumber\\   
  &&  + \frac{1} {(1-\xi_1)_+ (1-\xi_2)_+} \biggr ), 
 \nonumber\\
   {\mathcal B}_{1\sigma 1} (\xi_1,\xi_2) &=&  \xi_1(1-\xi_1) (\xi_1 +\txi2 -1), 
\nonumber\\
   {\mathcal B}_{1\sigma 2} (\xi_1,\xi_2) &=&  \delta (1-\xi_1) (\xi_2 -1) +\xi_1 (1-\xi_1-\txi2 ),
 \nonumber\\
   {\mathcal B}_{2\sigma 1} (\xi_1,\xi_2) &=& \xi_1^2  (-\xi_1 +2 \txi2 -1),
\nonumber\\
   {\mathcal B}_{2\sigma 2} (\xi_1,\xi_2) &=& 2 \delta (1-\xi_1)   \biggr (   
  (1-\xi_2)^2 L_1 (\xi_2)  +1  \biggr ) 
  +\frac{\xi_1^2}{(1-\xi_1)_+} (\xi_1 -2\txi2 +1),
\nonumber\\
   {\mathcal C}_{1\sigma } (\xi_1,\xi_2) &=& \delta (1-\xi_1) \frac{\xi_2}{(1-\xi_2)_+} 
         +\delta (1-\xi_2) \frac{\xi_1^2} {(1-\xi_1)_+}, 
\nonumber\\
    {\mathcal C}_{2\sigma } (\xi_1,\xi_2) &=& - 2 \xi_2 \delta (1-\xi_1) \biggr (  \frac{1}{(1-\xi_2)_+} +  L_1(\xi_2)  \biggr ) 
  -2\xi_1^2 \delta (1-\xi_2) \biggr (  \frac{1}{(1-\xi_1)_+} + L_2 (\xi_1)  \biggr )
\nonumber\\  
      &&   - \frac{\xi_1 (3\xi_1\txi2 -\xi_1 -\txi2 +1)}{(1-\xi_1)_+ (1-\xi_2)_+} , 
\nonumber\\
       {\mathcal D}_{1\sigma } (\xi_1,\xi_2) &=& \xi_1 (\xi_1-1) (\xi_1 +\tilde \xi_2 -1), 
\nonumber\\
    {\mathcal D}_{2\sigma } (\xi_1,\xi_2) &=& \xi_1^2 (\xi_1 -2 \tilde \xi_2 +1).                              
\end{eqnarray}                      
The variable $\tilde\xi_2$ is given by: 
\begin{equation} 
 \tilde \xi_2 = 1 -\xi_1 (1-\xi_2). 
\end{equation} 

\par


\vskip40pt
\begin{thebibliography}{99}


\bibitem{EFTE} A.V.~Efremov and O.V. Teryaev, Sov. J. Nucl. Phys. {\bf 36} 1982 142,
Phys. Lett. B150 (1985) 383.

\bibitem{QiuSt} J.W. Qiu and G. Sterman, Phys. Rev. Lett {\bf 67} (1991) 2264,
Nucl. Phys. B378 (1992) 52, Phys. Rev. D59 (1998) 014004.



\bibitem{DYSSA} N. Hammon, O. Teryaev and A. Schafer, Phys. Lett. B390 (1997) 409, arXiv:hep-ph/9611359, 

D. Boer, P. J. Mulders and O. V. Teryaev, Phys. Rev. D57 (1998) 3057,

D. Boer and P. J. Mulders, Nucl. Phys. B569 (2000)505,

D. Boer and J. W.  Qiu, Phys. Rev. D65 (2002) 034008.

\bibitem{ZM} J. Zhou and A. Metz, Phys. Rev. D86 (2012) 014001, e-Print: arXiv:1011.5871[hep-ph]. 

\bibitem{AT} I.V. Anikin and O.V. Teryaev, Phys. Lett. B690 (2010) 519,  
e-Print: arXiv:1003.1482 [hep-ph], Eur. Phys. C75 (2015) no 5, 184, e-Print: arXiv:1501.04536[hep-ph]. 

\bibitem{MaZh} J.P. Ma and G.P. Zhang, JHEP 1211 (2012) 156,  
e-Print: arXiv:1203.6415 [hep-ph]. 

\bibitem{CR} G. Re Calegari and P. G. Ratcliffe, Eur. Phys. J. C74 (2014) 2769, 
e-Print: arXiv:1307.5178 [hep-ph]. 

\bibitem{MaZh3}  J.P. Ma and G.P. Zhang, JHEP 1502 (2015) 163, 
e-Print: arXiv:1409.2938 [hep-ph].

\bibitem{JQVY} X. Ji, J.W. Qiu, W. Vogelsang and F. Yuan, Phys. Rev. Lett. {\bf 97} (2006) 082002, e-Print: hep-ph/0602239, 
 Phys. Rev. D73 (2006) 094017, e-Print: hep-ph/0604023 

\bibitem{KK1} K. Kanazawa and Y. Koike, Phys. Lett. B701 (2011) 576, e-Print:ar:Xiv1105.1036[hep-ph]. 

\bibitem{MSZ} J.P. Ma, H.Z. Sang and S.J. Zhu, Phys. Rev. D85 (2012) 114011, e-Print:arXiv:1111.3717. 


\bibitem{KY1} Y. Koike and S. Yoshida, Phys. Rev. D85 (2012) 034030, e-Print:arXiv:1112.1162[hep-ph].

\bibitem{MSS} J.P. Ma and H.Z. Sang,  JHEP 1104 (2011) 062, e-Print:arXiv:1102:1007[hep-ph]. 

\bibitem{MSC} J.P. Ma and H.Z. Sang, JHEP 0811 (2008) 090, e-Print:arXiv:0809.4811[hep-ph], H.G. Cao, J.P. Ma and H.Z. Sang, 
Commun. Theor. Phys. {\bf 53} (2010) 313, e-Print:arXiv:0901.2966[hep-ph].   

\bibitem{VoYu} W. Vogelsang and F. Yuan,  Phys.Rev. D79 (2009) 094010, e-Print: arXiv:0904.0410 [hep-ph].

\bibitem{KVX} Z.-B. Kang, I. Vitev and H.-X. Xing, Phys. Rev. D87 (2013) 034024, 
e-Print: arXiv:1212.1221[hep-ph]. 

\bibitem{DKPV} L.-Y. Dai, Z.B. Kang, A. Prokudin and I. Vitev, Phys. Rev. D92 (2015) no.11, 114024, e-Print:arXiv:1409.5851[hep-ph]. 

\bibitem{ShYo} S. Yoshida, Phys. Rev. D93 (2016) no.5, 054048, e-Print:arXiv:1601.07737[hep-ph]. 

\bibitem{G2} X. Ji and J. Osborne, Nucl. Phys. B608 (2001) 235, 

A.V. Belitsky, X.-D. Ji, W. Lu and J. Osborne, Phys. Rev. D63 (2001) 094012, e-Print:hep-ph/0007305, 

X.-D. Ji, W. Lu, J. Osborne and X.T. Song, Phys. Rev. D62 (2000) 094016, e-Print:hep-ph/0006121.  


\bibitem{CMW} A.P. Chen, J.P. Ma and G.P. Zhang, Phys. Lett. B754 (2016) 33, e-Print: arXiv:1505.03217[hep-ph]. 

\bibitem{EKT2} H. Eguchi, Y. Koike and K. Tanaka,  Nucl.Phys. B763 (2007) 198,  
e-Print: hep-ph/0610314. 

\bibitem{KoTa} Y. Koike and K. Tanaka, Phys.Lett. B646 (2007) 232-241, Erratum-ibid. B668 (2008) 458-459 
e-Print: hep-ph/0612117. 

\bibitem{KoTa2} Y. Koike and K. Tanaka, Phys. Rev. D76 (2007) 011502, e-Print: hep-ph/0703169.   

\bibitem{KTY} Y. Koike, K. Tanaka and S. Yoshida, Phys. Rev. D83 (2011) 114014, e-Print: arXiv:1104.0798[hep-ph]. 

\bibitem{JSPDF} J.C. Collins and D.E. Soper, Nucl. Phys. B194 (1982) 445. 

\bibitem{EKT} H. Eguchi, Y. Koike and K. Tanaka, Nucl. Phys. B752 (2006) 1, e-Print: hep-ph/0604003. 

\bibitem{JaJi} R.L. Jaffe, X.-D. Ji, Phys. Rev. Lett. 67 (1991) 552, Nucl. Phys. B375 (1992) 527-560.


\bibitem{JKT} J. Kodaira and K. Tanaka, Prog. Theor. Phys. 101 (1999) 191, e-Print: hep-ph/9812449. 

\bibitem{BD} A.V. Belitsky and D. Mueller,  Nucl. Phys. B503 (1997) 279,  e-Print: hep-ph/9702354.

\bibitem{ZYL1} J. Zhou, F. Yuan and Z.-T. Liang, Phys.Rev. D79 (2009) 114022 
e-Print: arXiv:0812.4484 [hep-ph].

\bibitem{Ji3G} X.D. Ji, Phys. Lett. B289 (1992) 137. 

\bibitem{KT1} Y. Koike and T. Tomita, Phys. Lett. B675 (2009) 181, e-Print:arXiv:0903.1923[hep-ph].


\bibitem{BKTY} H. Beppu, Y. Koike, K. Tanaka and Y. Yoshida, Phys. Rev. D82 (2010) 034005, e-Print: arXiv:1007.2034[hep-ph].

\bibitem{KangQiu1} Z.-B. Zhang and J.W. Qiu, Phys.Rev. D79 (2009) 016003,  
e-Print: arXiv:0811.3101 [hep-ph]. 

\bibitem{BMP}  V.M. Braun, A.N. Manashov and B. Pirnay, Phys. Rev. D80 (2009) 114002,
e-Print: arXiv:0909.3410 [hep-ph].

\bibitem{ZYL} J. Zhou, F. Yuan and Z.-T. Liang, Phys.Rev. D81 (2010) 054008, 
e-Print: arXiv:0909.2238 [hep-ph]. 

\bibitem{MW} J.P. Ma and Q. Wang, Phys. Lett. B715 (2012) 157, e-Print: arXiv:1205.0611[hep-ph]. 

\bibitem{SchZh} A. Schafer and J. Zhou, Phys. Rev. D85 (2012) 117501, e-Print:arXiv:1203.5293[hep-ph]. 

\bibitem{KangQiu2} Z.-B. Kang and J.W. Qiu, Phys. Lett. B713 (2012) 273,  
 e-Print: arXiv:1205.1019 [hep-ph]. 


\bibitem{Beli1} A.V. Belitsky, Phys. Lett.  B453  (1999)  59-72,  e-Print: hep-ph/9902361.

\bibitem{MWZh} J.P. Ma, Q. Wang and G.P. Zhang, Phys. Lett. B718 (2013) 1358, e-Print: arXiv:1210.1006. 

\bibitem{WVTE} W. Vogelsang, Phys. Rev. D57 (1998) 1856, e-Print: arXiv:hep-ph/9706511. 
\end{thebibliography}
\end{document}